\documentclass[10pt]{article}
\usepackage[margin=1in]{geometry}

\usepackage[english]{babel}

\usepackage[margin=1in]{geometry}
\usepackage{amsthm}

\usepackage{amsmath} % assumes amsmath package installed
\usepackage{float}
\usepackage{graphicx}
\usepackage{pdflscape}
\usepackage{amsfonts}
\usepackage{subcaption}
\usepackage{xcolor,colortbl}
\usepackage[hidelinks]{hyperref}
\hypersetup{
	colorlinks  = true,
	linkcolor  = {red!100!black},
	citecolor  = {blue!100!black},
	urlcolor     = {blue!100!black}
}
\usepackage{tabularx}
\usepackage{multirow}
\usepackage{algorithm}
\usepackage{algpseudocode}
\usepackage{booktabs}
\usepackage{nameref}
\usepackage{comment}
\usepackage{caption}
\usepackage[utf8]{inputenc}
\usepackage[english]{babel}
\newtheorem{Theorem}{Theorem}

\newtheorem{Remark}{Remark}
\newtheorem{Assumption}{Assumption}

\title{Data-Driven H-infinity Control with a Real-Time and Efficient Reinforcement Learning Algorithm: An Application to Autonomous Mobility-on-Demand Systems}
\author{Ali~Aalipour
        and~Alireza Khani
\thanks{
Ali Aalipour is with the Electrical and Computer Engineering department, University of Minnesota, MN, USA {\tt\small aalip002@umn.edu}.
\\
Alireza Khani is with the Civil, Environmental, and Geo-Engineering department, University of Minnesota, MN, USA {\tt\small akhani@umn.edu}.
}
}
\begin{document}
\maketitle

\begin{abstract}
Reinforcement learning (RL) is a class of artificial intelligence algorithms being used to design adaptive optimal controllers through online learning. This paper presents a model-free, real-time, data-efficient Q-learning-based algorithm to solve the H$_{\infty}$ control of linear discrete-time systems. The computational complexity is shown to reduce from $\mathcal{O}(\underline{q}^3)$ in the literature to $\mathcal{O}(\underline{q}^2)$ in the proposed algorithm, where $\underline{q}$ is quadratic in the sum of the size of state variables, control inputs, and disturbance. 
An adaptive optimal controller is designed and the parameters of the action and critic networks are learned online without the knowledge of the system dynamics, making the proposed algorithm completely model-free.
Also, a sufficient probing noise is only needed in the first iteration and does not affect the proposed algorithm. With no need for an initial stabilizing policy, the algorithm converges to the closed-form solution obtained by solving the Riccati equation. 
A simulation study is performed by applying the proposed algorithm to real-time control of an autonomous mobility-on-demand (AMoD) system for a real-world case study to evaluate the effectiveness of the proposed algorithm.
% In this paper, we present a model-free, real-time, data-efficient Q-learning-based algorithm to solve the H$_{\infty}$ control of linear discrete-time systems, in particular, to optimize vehicle scheduling and rebalancing in an autonomous mobility-on-demand (AMoD) system that can be modeled as a linear discrete-time system. To fulfill customer demands, rebalancing is crucial to ensure that vehicles are distributed properly which means dispatching available empty vehicles to areas undersupplied areas. We show that computational complexity reduces from $\mathcal{O}(\underline{q}^3)$ to $\mathcal{O}(\underline{q}^2)$ compared to the available algorithm in the literature, where $\underline{q}$ is quadratic in the sum of the size of state variables, control inputs, and disturbance. 
% An adaptive optimal controller is designed and the action and critic networks's parameters are learned online without the knowledge of the system dynamics,
% making the proposed algorithm completely model-free.
% Also, a sufficient probing noise is only needed in the first iteration and does not affect the proposed algorithm. With no need for an initial stabilizing policy, the algorithm converges to the closed-form solution obtained by solving the Riccati equation. 
% A simulation study is performed by applying the proposed algorithm to an autonomous mobility-on-demand (AMoD) system in real-time for a real-world case study to evaluate the effectiveness of the proposed algorithm.
\end{abstract}

\section{Introduction}

Nowadays, people are able to use mobility-on-demand (MoD) services to travel and share vehicles with other people by sending requests through mobile devices. MoD can be replaced by AMoD due to the lower costs of autonomous vehicle (AV) operations \cite{nguyen2023examining}. The cooperative nature of AVs is in contrast with selfish taxi drivers seeking to maximize their profits. By optimizing routing, rebalancing, charging schedules, etc., central coordination can minimize externalities in AMoD systems. Further, customers do not have to drive, which allows them to save time on commuting. Various companies started to develop this technology in response to such promising benefits.

Fleet management studies focus on optimizing vehicle routing to rebalance empty vehicles and serve customers in the network. They aim to reduce the operational costs and the waiting times of customers. The AMoD system avoids the costs of rebalancing drivers to drive vehicles from oversupplied origins to undersupplied origins. Moreover, similar to the current car-sharing companies such as Car2go, AMoD provides excellent convenience for one-way trips since users do not have to return vehicles to the origin of the trip. AMoD services, therefore, provide opportunities for efficient fleet management. 

If the AMoD framework is not appropriately controlled, it can run into an imbalanced system, i.e., oversupplying the stations frequently opted as destinations. In contrast, regions with a high number of originating trips are undersupplied. To circumvent this issue, a rebalancing strategy is needed in moving vehicles to high-demand stations. Aiming for the rebalancing strategy, we need a model to capture the dynamics of the AMoD system.

\begin{comment}
 Model-Free RL: Explicit control models offer some advantages, in most real-world applications the relevant dynamics are incredibly complex, and deriving exact models is not possible. Moreover, model-based methods may require strong assumptions and relaxations, e.g. constant arrival rates that are difficult to
validate in real-world applications, and may scale poorly, e.g., due to integer constraints. As a result, model-free methods based on RL for addressing vehicle dispatch and rebalancing
have started to gain popularity in the literature. The most significant advantage of model-free RL compared to model-based approaches is that a forward simulator suffices to compute a stochastic control policy without relying on an explicit
mathematical model of the system dynamics or exogenous factors. Furthermore, the use of deep neural networks makes it possible to evaluate the policy in real-time because computing actions is equivalent to making one forward pass, and this
step can be further accelerated via the use of specialized hardware such as a Graphical Processing Unit (GPU). Well-known challenges for model-free RL methods include high
sample complexity, the need for a trade-off between policy
exploration and exploitation, problem-dependent reward shaping, and the determination of appropriate neural architecture
for the policy network. As is also the case with model-based
methods, differing formulations of the state and action spaces
can have a large impact on control performance.   
\end{comment}

\subsection{Literature Review}
At a high level, approaches to tackling operational AMoD issues may be divided into two categories: model-free and model-based.
\subsubsection{Model-free}
Model-free techniques for AMoD fleet rebalancing can be characterized as centralized or decentralized. A centralized agent rebalances cars in order to optimize specific objectives, such as travel time. 
The authors in \cite{turan2020dynamic} present a RL approach that uses a dynamic pricing AMoD framework aiming to maximize the profit and rebalance the fleet. 
On the other hand, \cite{skor2021modular} emphasizes customer waiting time minimization. A RL approach for the taxi dispatching and rebalancing problem is introduced in \cite{gao2018optimize} to maximize taxi driver long-term revenues utilizing Q-learning and discrete states and actions on a grid-shaped map. A double deep Q-learning architecture is proposed in \cite{guo2022deep} for vehicle routing in a ride-sharing AMoD system, where idle cars are rebalanced to fulfill future demands. \cite{fluri2019learning} employs RL and mixed integer linear programming (MILP) for fleet rebalancing and uses hierarchical binary partitioning and tabular Q-learning for RL.

Decentralized approaches, as contrasted with centralized methods, enable each vehicle to act as its own agent and be trained in either a cooperative or competitive fashion. In \cite{al2019deeppool}, a ride-sharing architecture for vehicles utilizing a deep Q-network to learn the optimal policies is proposed for individual vehicles in a distributed and uncoordinated fashion. Passenger satisfaction and vehicle utilization are the two most important objectives of the framework. The authors in \cite{lin2018efficient} employ multi-agent RL, where each vehicle functions as an individual agent. Similarly, \cite{gueriau2018sAMoD} provides a dynamic ride-sharing system in which both passenger assignments and fleet rebalancing are learned and performed by individual agents using multi-agent RL. Furthermore, \cite{wen2017rebalancing} addresses rebalancing idle vehicles by developing a deep Q-learning approach.
\subsubsection{Model-based}
Model-based AMoD techniques attribute an explicit model to system dynamics and utilize it to determine optimal decisions. Despite their complexity, they are powerful and allow us to examine the model's properties, including convergence. Numerous studies proposed and developed system models including queuing \cite{zhang2016models,volkov2012markov}, fluidic \cite{pavone2012robotic}, network flow \cite{zhang2016control,rossi2018routing,carron2019scalable}, and data-driven \cite{lei2020efficient} approaches. Further classifications of model-based approaches include mathematical optimization and simulation-based methods. Various studies have tackled the rebalancing of vehicle fleets as a complex optimization problem. Combining the model predictive control (MPC) algorithms with the network flow model offers an efficient tool for expressing complex constraints. For example, \cite{iglesias2018data} implements an MPC algorithm leveraging historical data and neural networks to develop a model for short-term demand forecasts to address the dispatching and rebalancing problem. Moreover, to improve social welfare, \cite{tsao2019model} suggests a real-time MPC framework that optimizes a weighted combination of vehicle mileage and passenger travel time.
% \cite{tsao2019model} proposes a real-time MPC framework to promote social welfare by optimizing the routes of both customer-carrying and empty vehicles, i.e., a weighted combination of vehicle mileage and passenger travel time. 
In addition, a scalable MPC control has been developed in \cite{carron2019scalable} to keep the system balanced.

Reinforcement learning (RL) is one of the three fundamental machine learning paradigms, alongside supervised learning and unsupervised learning, which has a long history \cite{sutton2018reinforcement}. The RL discipline has also been reinvented by recent developments in machine learning (ML), particularly employing deep networks. The dynamical system's model is typically unknown in RL settings, and the ideal controller is discovered by engagement with the environment. It is fundamental for the RL algorithms to deliver assured stability and performance as the range of RL extends to more difficult tasks. Due to deep networks' inherent complexity and the intricacy of the tasks, we are still a long way from being able to analyze RL algorithms. This encourages thinking about a case study that is simplified and allows for analysis. There are well-known challenges associated with model-free RL algorithms. Examples of these challenges include the necessity of a trade-off between policy exploitation and exploration, problem-dependent reward shaping, and the design of an appropriate neural architecture for the policy network. In addition, the majority of them are neither theoretically tractable nor can their convergence be investigated. 
% Additionally, the performance of the controller may be significantly impacted by various formulations of the state and action spaces.

H$_{\infty}$ problem is a classical control problem where the dynamical system follows linear dynamics and the cost function to be minimized is quadratic. It is a robust control method that is implemented to attenuate the effects of disturbances on the performance of dynamical systems. It is a great benchmark for studying since the closed-form solution for H$_{\infty}$ is available. Moreover, it is theoretically tractable in comparison to the RL algorithms. 

As a result of the aforementioned factors, the linear quadratic (LQ) problem has received greater attention from the RL community \cite{abbasi2019model,dean2020sample,tu2018least}, see also \cite{matni2019self} for a thorough overview of RL methods and their properties for the LQ problems.
% For example, \cite{kiumarsi2017h} studies the Linear Quadratic Regulator (LQR) control problems in deterministic setups via LSTD.
In addition, the convergence of policy gradient methods for the linear quadratic regulator (LQR) problem is shown in \cite{fazel2018global}. RL also has been applied for solving optimal control problems in an uncertain environment,\cite{al2007model,wu2012neural,kiumarsi2017h}. Inherently, the Q-learning algorithm does not eliminate the impacts of the probing noise, which is employed to excite the system, in the Bellman equation when evaluating the value function. The algorithm's convergence may be impacted, and this may lead to bias. In \cite{kiumarsi2017h}, two separate policies are used to update the algorithm to cancel the effects of probing noise. However, there should be enough generated data for each iteration to estimate the policies. 
% Moreover, the disturbance input must be updated in the prescribed optimal fashion (18) and applied to system dynamics to collect data. However, in practical applications, the disturbance is independent and cannot be specified. In Section 3, it is shown that the proposed method fixes this issue.

\subsection{Contributions}
In this paper, we propose a RL algorithm to solve the H$_{\infty}$ control of linear discrete-time systems.
It is model-free, real-time, and data-efficient, i.e., using a single data, the parameters of the actor and critic networks are updated. This feature results in reducing the order of computational complexity to square ($\mathcal{O}(\underline{q}^2)$) where $\underline{q}$ is the number of parameters being estimated compared to the cube order ($\mathcal{O}(\underline{q}^3)$) in the state-of-the-art algorithms in the literature (e.g., \cite{al2007model,kiumarsi2017h}). 
% We show that computational complexity decreases to $\mathcal{O}(\underline{q}^2)$ compared to the available algorithm in the literature.
This RL algorithm does not suffer from bias if probing noise is used. Moreover, a sufficient amount of probing noise is only needed in the first iteration, i.e., the policy used to generate data, called the behavior policy is different only in the first iteration than the policy that is being evaluated and improved, called the estimation policy or target policy.
The convergence of the proposed algorithm is shown. Moreover, we apply the proposed algorithm to an AMoD system which can be modeled as an H$_{\infty}$ control of linear discrete-time systems.
In summary, the contributions of this paper can be expressed as follows:
\begin{enumerate}
    \item Proposed a model-free, real-time, and data-efficient algorithm to solve the H$_{\infty}$ control of linear discrete-time systems.
\item Reduced the order of computational complexity form cube ($\mathcal{O}(\underline{q}^3)$) in the state-of-the-art algorithms in the literature to square ($\mathcal{O}(\underline{q}^2)$).
\item Discussed the properties of the proposed algorithm and proved its convergence.
\item Applied the proposed algorithm to an AMoD system which can be modeled as an H$_{\infty}$ control of linear discrete-time systems.
\end{enumerate}

\subsection{Organization}
The remainder of the paper is organized as follows. 
Section \ref{sec: AMoD system modeling} presents the problem formulation and model for the AMoD system.
In Section \ref{sec: H inf section}, the discrete-time H$_{\infty}$ control problem is formulated. This section is concluded by implementing the  Q-learning algorithm.
In Section \ref{sec: Implement RL Alg},
the online implementation of the proposed algorithm and its properties are analyzed. Besides, the convergence of the proposed algorithm is proved. We present the results of the numerical case study example in Section \ref{sec: results}. Finally, the paper is concluded in Section \ref{sec: conclusion}.
\subsection{Notation}
The symbols ${\bf 1}_m$ and ${\bf 0}_n$ denote column vectors of dimension $m$ and $n$ with all entries equal to $1$ and $0$, respectively. Given a vector ${p} \in \mathbb{R}^n$, we define $\tilde {P} = {\rm diag} \left ( { p}\right) \in \mathbb{R}^{n\times n}$ as a diagonal matrix with the elements of the vector $p$ on the diagonal. ${\rm vecs}(P)=\begin{bmatrix}
p_{11},...,p_{1n},p_{22},...,p_{2n},...,p_{nn}\end{bmatrix}^T$ is the vectorization of the upper-triangular part of a symmetric matrix $P \in \mathbb{R}^{n\times n}$, and  
${\rm vecv}(v)=\begin{bmatrix}
v_1^2,2v_1v_2,...,2v_1v_n,v_2^2,...,2v_2v_n,... v_n^2\end{bmatrix}^T$ is the quadratic vector of the vector $v \in \mathbb{R}^{n}$.

\subsection{Preliminaries}
Consider a directed graph ${G}\left(N,A\right)$ where $N=\left\{1, \dots, n\right\}$ is the set of nodes and $A=\left\{1, \dots, m\right\}$ is the set of links. Let $E_{\rm in}$ and $E_{\rm out}\in \left\{0,1\right\}^{n\times m}$ be the in-neighbors and out-neighbors matrices.
% If a link enters a node, the associated entry is one in $E_{\rm in}$. Similarly, if a link exits a node, the associated entry is equal to one in the matrix $E_{\rm out}$. 
The incidence matrix $E \in \left\{ -1,0,1\right\}^{n \times m}$ can be derived by $E = E_{\rm in} -E_{\rm out}$.

\section{Discrete-time (DT) H$_{\infty}$ Control Problem}
\label{sec: H inf section}
Consider the following linear discrete-time system
\begin{align}
x_{t+1} = {\mathcal A}x_t + {\mathcal B} v_t + {\mathcal L} d_t,
\label{eq: mainSYS}
\end{align}
where $x_t \in {\mathbb R}^{m_1}$ is the system state, $v_t \in {\mathbb R}^{m_2}$ is the control input, and $d_t \in {\mathbb R}^{m_3}$ is the external disturbance input.
\begin{Assumption}
The pair $({\mathcal A}, {\mathcal B})$ is stabilizable, i.e., all uncontrollable modes are asymptotically stable.
\end{Assumption}
% \label{sec: Q Learning}
% In this section, 
We consider the standard Q-learning algorithm and discuss its properties. Since system identification is not going to be performed to estimate the parameters of systems, we use the following objective function
\begin{align}
\label{eq: maincost}
\mathcal{J}(x_t,v_t,d_t)
&= \sum_{i=t}^{\infty} r(x_i,v_i,d_i)
\end{align}
where $$r(x_i,v_i,d_i) = x_i^T R_x x_i + v_i^T R_v v_i -\gamma^2 d_i^Td_i,$$
for a prescribed fixed value of $\gamma$. 
Matrices $R_x$ and $R_v$ are positive semidefinite (PSD) and positive definite (PD), respectively. In the H$_\infty$ control problem, $\gamma$ is an upper bound on the desired $L_2$ gain disturbance attenuation \cite{baar1995if}. Note that the formulation we used is similar to min-max LQ in \cite{rantzer2021minimax} and \cite{zhang2019policy}. In particular, the authors in \cite{zhang2019policy}
consider a nonconvex-nonconcave saddle-point problem in the policy space and show that despite its non-convexity and
non-concavity, zero-sum LQ games have the property that the stationary point of the objective function with respect to the linear feedback control policies constitutes the Nash equilibrium (NE) of the game. In the zero-sum game LQ problem, it is desired to find the optimal control $v^*_t$ and the worst-case disturbance $d^*_t$. Note that functions in $L_2\left.[0,\infty\right.)$ represent the signals having finite energy over infinite interval $\left.[0,\infty\right.)$. That is,
% \begin{equation}
$\sum\limits_{t = 0}^\infty {d_t^Td_t} < \infty$.
% \end{equation}
Moreover, using \eqref{eq: maincost} and given some fixed policy for
an admissible control policy ${v_t} = K_v{x_t}$ and a disturbance policy ${d_t} = K_d{x_t}$ the value function is defined as
\begin{equation}
{V}(x_{t},K_v,K_d) = \sum\limits_{i = t}^\infty {r(x_i,K_v x_i,K_d x_i)},
\label{eq: value function}
\end{equation}
and the Bellman equation reads
\begin{equation}
{V}(x_{t},K_v,K_d)= {r(x_t,K_v x_t,K_d x_t)} + {V}(x_{t+1},K_v,K_d).
\label{eq: RL bellman}
\end{equation}
% Moreover, given some control policy $v_t = \mu_v(x_t)$ and a disturbance policy $d_t = \mu_d(x_t)$, define the $Q$ function
% for that policy as
% \begin{equation}
% Q^\mu (x_t,v_t,d_t) = x_t^T R_x x_t + v_t^T R_v v_t -\gamma^2 d_t^Td_t + V^\mu(x_{t+1}).
% \label{eq: RL bellman Q}
% \end{equation}
Since $V(x_t, K_v,K_d) = Q(x_t,K_v x_t,K_d d_t)$, the Bellman equation under the policy gains $K_v$ and $K_d$ can be rewritten as follows:
\begin{align}
Q(x_t,v_t,d_t) =& {r(x_t,v_t,d_t)} + V(x_{t+1}, K_v,K_d),
\label{eq: RL bellman Q}
\end{align}
and the Bellman optimality equation for the Q-function under the optimal policy gains $K_v^{\star}$ and $K_d^{\star}$ is
\begin{align}
Q^{\star} (x_t,v_t,d_t) = & {r(x_t,v_t,d_t)} + Q^{\star} (x_{t+1},K_v^{\star}x_{t+1},K_v^{\star} x_{t+1}).
\end{align}
% or 
% \begin{align}
% Q^* (x_t,v_t,d_t) =& x_t^T R_x x_t + v_t^T R_v v_t -\gamma^2 d_t^Td_t \notag \\
% &+\arg\min_{v_{t+1}}\max_{d_{t+1}}Q^*(x_{t+1},v_{t+1},d_{t+1})
% \end{align}
\subsection{Derivation of Q-learning Algorithm}
% The value iteration algorithm (\cite{sutton2018reinforcement,lewis2012reinforcement}) is applied to the Bellman equation \eqref{eq: RL bellman Q}.
We use the Q-function to develop a Q-learning algorithm (\cite{sutton2018reinforcement,lewis2012reinforcement}) to solve for the DT H$_{\infty}$ Control Problem using the Bellman equation \eqref{eq: RL bellman Q}.
This routine is an actor-critic class of reinforcement learning, where the critic agent evaluates the current control policy using methods based on the policy evaluation. After this evaluation is completed, the action is updated by an actor agent based on the policy improvement. The learning process starts with an initial Q-function $Q^0(x,v,d) = 0$ in the Q-learning that is not necessarily optimal, and then derives $Q^1(x,v,d)$ by solving Eq. \eqref{eq: iterative Q Bellman} with $i = 0$.
\subsubsection{Policy evaluation}
We evaluate the policy by using $Q$-function in \eqref{eq: iterative Q Bellman}.\\
\begin{align}
Q^{i+1} (x_t,v_t,d_t) =& {r(x_t,v_t,d_t)}+ Q^{i} (x_{t+1},K_v^ix_{t+1},K_v^i x_{t+1}).
\label{eq: iterative Q Bellman}
\end{align}

\subsubsection{Policy improvement} 
The control and disturbance policies will be improved as follows:
\begin{align*}
K_v^{i+1}= & \arg\min_{K_v} Q^{i+1}(x_t,v_t,d_t)
\\
K_d^{i+1}= & \arg\max_{K_d} Q^{i+1}(x_t,v_t,d_t).
\end{align*}
Let $z_t = [x_t^T,v_t^T,d_t^T]^T$
% $z_{t+1} = [x_{t+1}^T,(K_v^ix_{t+1})^T,(K_d^ix_{t+1})^T]^T$
and $P^i = \left [ \begin{matrix}
 I&{K_v^i}^T&{K_d^i}^T
\end{matrix} \right ]S^i\begin{bmatrix}
 I&{K_v^i}^T&{K_d^i}^T
\end{bmatrix}^T.$
Given a linear system, linear policies, and quadratic cost, we can assume the quality function (Q-function) is quadratic in the state, control, and disturbance so that
\begin{equation}
Q^{i+1}(z_t) =z_t^TS^{i+1}z_t.
\label{eq: Q est}
\end{equation}
% and equivalent to the model-based Bellman equation.
Applying \eqref{eq: Q est} in \eqref{eq: iterative Q Bellman}, the Lyapunov equation yields
\begin{equation}
z_t^TS^{i+1}z_t= {r(x_t,v_t,d_t)} + x_{t+1}^TP^{i}x_{t+1}.
\label{eq: Lyapunov Equation Q}
\end{equation}
Replacing the dynamics \eqref{eq: mainSYS} in 
\eqref{eq: Lyapunov Equation Q}, we have:
\begin{align*}
z_t^T  S^{i+1}z_t =&  {x_t^T R_x x_t + v_t^T R_v v_t -\gamma^2 d_t^Td_t}  +({\mathcal A}x_t+{\mathcal B}v_t+{\mathcal L}d_t)^TP^i({\mathcal A}x_t+{\mathcal B}v_t+{\mathcal L}d_t)
\\ 
= & \begin{bmatrix}
x_t^T & v_t^T& d_t^T
\end{bmatrix}
 \begin{bmatrix}
R_x+{\mathcal A}^TP^i{\mathcal A} & {\mathcal A}^TP^i{\mathcal B} & {\mathcal A}^TP^i{\mathcal L}\\ 
{\mathcal B}^TP^i{\mathcal A} & R_v+{\mathcal B}^TP^i{\mathcal B} & {\mathcal B}^TP^i{\mathcal L}\\
{\mathcal L}^TP^i{\mathcal A} & {\mathcal L}^TP^i{\mathcal B} & {\mathcal L}^TP^i{\mathcal L}-\gamma^2I
\end{bmatrix}\begin{bmatrix}
x_t \\ v_t\\ d_t
\end{bmatrix}\\
= & z_t^T\begin{bmatrix}
R_x+{\mathcal A}^TP^i{\mathcal A} & {\mathcal A}^TP^i{\mathcal B} & {\mathcal A}^TP^i{\mathcal L}\\ 
{\mathcal B}^TP^i{\mathcal A} & R_v+{\mathcal B}^TP^i{\mathcal B} & {\mathcal B}^TP^i{\mathcal L}\\
{\mathcal L}^TP^i{\mathcal A} & {\mathcal L}^TP^i{\mathcal B} & {\mathcal L}^TP^i{\mathcal L}-\gamma^2I
\end{bmatrix}z_t.
\end{align*}

Let us partition matrix $S^{i+1}$ as 
\begin{align}
S^{i+1} = \begin{bmatrix}
S^{i+1}_{xx}&S^{i+1}_{xv} &S^{i+1}_{xd}\\
S^{i+1}_{vx}&S^{i+1}_{vv} & S^{i+1}_{vd}\\
S^{i+1}_{dx}&S^{i+1}_{dv} & S^{i+1}_{dd}
\end{bmatrix}.    
\end{align}
Optimizing $Q^{i+1}(z_t)$ over $v_t$ and $d_t$ results in 
\begin{align*}
& v_t = - {S^{i+1}_{vv}}^{-1} (S^{i+1}_{vd} d_t + S^{i+1}_{vx}x_t),\\
& d_t = - {S^{i+1}_{dd}}^{-1} (S^{i+1}_{dv} v_t + S^{i+1}_{dx}x_t).
\end{align*} 
Substituting $v_t$ in $d_t$ and vice versa yields the equations $v_t^{i+1} =K_v^{i+1} x_{t}$ and $d_t^{i+1} =K_d^{i+1} x_{t}$
where 
\begin{subequations}
\begin{align}
K_v^{i+1} =& \left( S^{i+1}_{vv} - S^{i+1}_{vd}{S^{i+1}_{dd}}^{-1} S^{i+1}_{dv} \right)^{-1} \left( S^{i+1}_{vd}{S^{i+1}_{dd}}^{-1} S^{i+1}_{dx} - S^{i+1}_{vx}\right),
    \label{eq: opt cont gain}\\
K_d^{i+1} =& \left( S^{i+1}_{dd} - S^{i+1}_{dv}{S^{i+1}_{vv}}^{-1} S^{i+1}_{vd} \right)^{-1} \left( S^{i+1}_{dv}{S^{i+1}_{vv}}^{-1} S^{i+1}_{vx} - S^{i+1}_{dx}\right).
\label{eq: opt dist gain}
\end{align}
\end{subequations}
Using \eqref{eq: Q est} and applying the above result in \eqref{eq: Lyapunov Equation Q},
the following recursion can be concluded:
\begin{align}
&S^{i+1}= 
\underbrace{\begin{bmatrix}
R_x & 0 & 0\\
0 & R_v & 0\\
0 & 0 & -\gamma^2I
\end{bmatrix}}_{G} +\begin{bmatrix}
{\mathcal A}^T\\ 
{\mathcal B}^T\\
{\mathcal L}^T
\end{bmatrix}
\begin{bmatrix}
I & {K_v^i}^T&{K_d^i}^T
\end{bmatrix}S^i
\begin{bmatrix}
I \\K_v^i\\K_d^i
\end{bmatrix}
\begin{bmatrix}
{\mathcal A} & {\mathcal B} & {\mathcal L}
\end{bmatrix}.
\label{eq: it on S}
\end{align}

Given $$P^i = \left [ \begin{matrix}
 I&{K_v^i}^T&{K_d^i}^T
\end{matrix} \right ]S^i\begin{bmatrix}
 I&{K_v^i}^T&{K_d^i}^T
\end{bmatrix}^T,$$
the following equation can be concluded:
$$P^{i+1} = \left [ \begin{matrix}
 I&{K_v^{i+1}}^T&{K_d^{i+1}}^T
\end{matrix} \right ]S^{{i+1}}\begin{bmatrix}
 I&{K_v^{i+1}}^T&{K_d^{i+1}}^T
\end{bmatrix}^T.$$
Substituting \eqref{eq: it on S}, \eqref{eq: opt cont gain}, and \eqref{eq: opt dist gain}, one can obtain:
\begin{align}
P&^{i+1} = R_x+{\mathcal A}^TP^i{\mathcal A} -\begin{bmatrix}
{\mathcal A}^TP^i{\mathcal B} & {\mathcal A}^TP^i{\mathcal L}
\end{bmatrix}
\begin{bmatrix}
 R_v+B^TP^i{\mathcal B} & {\mathcal B}^TP^i{\mathcal L}\\
 {\mathcal L}^TP^i{\mathcal B} & {\mathcal L}^TP^i{\mathcal L}-\gamma^2I
\end{bmatrix}^{-1}
\begin{bmatrix}
{\mathcal B}^TP^i{\mathcal A} \\ {\mathcal L}^TP^i{\mathcal A}
\end{bmatrix}.
\label{eq: ricatti recursion}
\end{align}
Equation \eqref{eq: ricatti recursion} is called Lyapunov Recursion.

In summary, we evaluate the policy gains $K_v$ and $K_d$ by finding the quadratic kernel $S$ of the $Q$-function using \eqref{eq: it on S} and then improved policy gains are given by \eqref{eq: opt cont gain} and \eqref{eq: opt dist gain}.\\

\section{Online Implementation of the Proposed Algorithm}
\label{sec: Implement RL Alg}
In this section, we discuss the online implementation of the proposed algorithm and prove its convergence. Algorithm \ref{Alg: Alg1} summarizes the steps of the proposed algorithm for the H$_{\infty}$ problem \eqref{eq: mainSYS}.
\begin{algorithm}
\caption{}
\label{Alg: Alg1}
\begin{algorithmic}[1]
\State {\bf Initialization:} $i=0$, Any {\it \bf arbitrary} policy gain $K_v^0$, $K_d^0$, and $S = {\bf 0}$
\For{$ \tau = -(q-1),\cdots,0$}

\State Sample $\lambda \sim \mathcal{N}\left(0,W_{\lambda}\right)$ and set $v = K_v^0x+\lambda$.

\State Take $v$ and $d$ and observe $x_+$.

% Update Z as $Z = \{Z,(x,v,d,x_+,K_v^0x,K_d^0x)\}$.
\EndFor
\State Estimate $S^{1}$ by \eqref{eq: main estimation optimization}

\State Improve the policies $K_v^{1}$ and $K_d^{1}$ by \eqref{eq: opt cont gain} and \eqref{eq: opt dist gain}.
% \State {\bf Step 2} Policy evaluation and policy improvement
\While { $\|S^{i+1} -S^i\|_2 >\epsilon$}
% {$K_v^i,K_d^i$, $i = 1,\cdots,N$
% }

\State Take $K_v^{i}$ and $K_d^{i}$ and observe $x_{i+1}$.

% $Z = \{Z,(x_{i},K_v^{i}x_{i},K_d^{i}x_{i},x_{i+1},K_v^{i}x_{i+1},K_d^{i}x_{i+1})\}$.

% \State {\bf{Policy Evaluation}}
\State Estimate $S^{i+1}$ by \eqref{eq: main estimation optimization}.

% \State {\bf{Policy Improvement}}
\State Improve the policies by \eqref{eq: opt cont gain} and \eqref{eq: opt dist gain}.

\State $i = i+1$.
\EndWhile
\end{algorithmic}
\end{algorithm}
We will parameterize the Q-function in \eqref{eq: iterative Q Bellman} so that we can separate the unknown
matrix $S$. Using parameterization and defining $s = {\rm vecs}(S)$, $p = {\rm vecs}(P)$, ${{z}_t}=
 \begin{bmatrix}
 {x_t^T},{v_t^T}, {d_t^T}
\end{bmatrix}^T$, ${\bf \phi}_t(K_v^i) = [ x_t^T,(K_v^ix_t)^T,{d}_t^T]^T$, and ${\bf \phi}_t(K_v^i, K_d^i) = [ x_t^T,(K_v^ix_t)^T,(K_d^ix_t)^T]^T$, we have the below equation:
\begin{equation}
\label{parameterize}
{\rm vecv}(
z_t){s_{i + 1}} =r\left(x_t,v_t,d_t\right) + {\rm vecv}({{\phi}_{t+1}} (K_v^i, K_d^i)){s_i}.
\end{equation}
To find the optimal policy in each iteration, we need to solve the following least square (LS) problem:
\begin{align}
s_{i + 1} = \mathop {\min {\mkern 1mu} }\limits_{s}\quad \left \|{\Psi _{i}}{s} -\Gamma_i   \right \|_2^2,
 \label{eq: main estimation optimization}
\end{align}
where:
\begin{itemize}
\item  $\xi  = {\rm vecs}\left( G \right).$
\item $\Gamma_{i} = {\Psi_{i}}\xi + \Phi_{i}{s^{i}} = \left [ \begin{matrix}
\Gamma_{i-1}^T,\gamma_{i}^T
\end{matrix} \right ]^T$
where $\gamma_{i} = {\rm vecv}(\phi_i(K_v^i))\xi
+{\rm vecv}(\phi_{i+1}(K_v^i, K_d^i))s^{i}
$. 
\item $\Phi_{i}{s_{i}}$ can be written as $X^+_{i}p_{i}$ where $X^+_{i} =\left [ \begin{matrix}
{X^+_{i-1}}^T, {\rm vecv}(x_{i+1})^T
\end{matrix} \right]^T$.
\item $\Psi_{i} = \left[\begin{matrix}
{\Psi_{i-1}}^T,
{\rm vecv}(\phi_i(K_v^i))^T\end{matrix} \right]^T$
and  \\$\Phi_{i} = \begin{bmatrix}
\Phi_{i-1}^T,{\rm vecv}(\phi_{i+1}(K_v^i,K_d^i))^T
\end{bmatrix}^T$, for $i = 1,2,\cdots$.
\item The initial values are given as $\Psi_0  =  \begin{bmatrix}
{\rm vecv}(z_{-(q-1)})^T ,
{\rm vecv}(z_{-(q-2)})^T,
 \cdots,
{\rm vecv}(z_{0})^T
\end{bmatrix}^T$,
\\ 
${\Phi_0} = \begin{bmatrix}
{\rm vecv}(\phi_{-(q-2)}(K_v^0,K_d^0))^T,&\cdots&,
{\rm vecv}(\phi_1(K_v^0,K_d^0))^T\end{bmatrix}^T$,
$\Gamma_0 = \Psi_0\xi + \Phi_0{s^0}$, and \\$X^+_0 = \left [ \begin{matrix}
{\rm vecv}(x_{-(q-2)}))^T,&\cdots&
,{\rm vecv}(x_{1})^T\\ 
\end{matrix} \right]^T
$.
\end{itemize}

% $\Gamma_{i} = {\Psi_{i}}\xi + \Phi_{i}{s^{i}} = \left [ \begin{matrix}
% \Gamma_{i-1}^T,\gamma_{i}^T
% \end{matrix} \right ]^T$
% where $\gamma_{i} = {\rm vecv}(\phi_i(K^i))\xi
% +{\rm vecv}(\phi_{i+1}(K^i))s^{i}
% $. $\Phi_{i}{s_{i}}$ can be written as $X^+_{i}p_{i}$ where $X^+_{i} =\left [ \begin{matrix}
% {X^+_{i-1}}^T, {\rm vecv}(x_{i+1})^T
% \end{matrix} \right]^T$.
% Also,$\Psi_{i} = \left[\begin{matrix}
% {\Psi_{i-1}}^T,
% {\rm vecv}(\phi_i(K^i))^T\end{matrix} \right]^T$
% and  $\Phi_{i} = \begin{bmatrix}
% \Phi_{i-1}^T,{\rm vecv}(\phi_{i+1}(K^i))^T
% \end{bmatrix}^T$, for $i = 1,2,\dots$.
% And, the initial values are given as
% $\Psi_0  =  \begin{bmatrix}
% {\rm vecv}(z_{-(q-1)})^T ,
% {\rm vecv}(z_{-(q-2)})^T,
%  \cdots,
% {\rm vecv}(z_{0})^T
% \end{bmatrix}^T$,
% ${\Phi_0} = \begin{bmatrix}
% {\rm vecv}(\phi_{-(q-2)}(K^0))^T,&\cdots&,
% {\rm vecv}(\phi_1(K^0))^T\end{bmatrix}^T$,
% $\Gamma_0 = \Psi_0\xi + \Phi_0{s^0}$, and $X^+_0 = \left [ \begin{matrix}
% {\rm vecv}(x_{-(q-2)}))^T,&\cdots&
% ,{\rm vecv}(x_{1})^T\\ 
% \end{matrix} \right]^T
% $.
% Also, let's denote $$G=\begin{bmatrix}
% R_x&0&0\\
% 0&R_v&0\\
% 0&0&-\gamma^2I
% \end{bmatrix},$$ and $\xi  = {\rm vecs}\left( G \right)$.
Equation \eqref{parameterize} is used in the policy evaluation step to solve for the unknown vector $s$ in the least-squares sense by collecting $q \ge \underline{q}$ data samples of $x$, $v$, and $d$, where $\underline{q} = {{(m_1 + m_2+m_3)(m_1 + m_2+m_3+ 1)}/ 2}$.
% Note that these $q$ data samples refer to different points in time, $k$, which are used to form the data matrices ${\Psi}\in{{\mathbb R}^{q \times {{(m + n)(m + n + 1)} \mathord{\left/
%  {\vphantom {{(m + n)(m + n + 1)} 2}} \right.
%  \kern-\nulldelimiterspace} 2}}}$ and $\Gamma  \in {{\mathbb R}^{q \times 1}}$ defined by,....
It should be noted that $v_t$ and $d_t$ are linearly dependent on $x_t$ which means that $ {{\Psi ^T}\Psi } $ is not invertible. To resolve this issue, excitation noise is added in $v_t$ and $d_t$ in only the first iteration such that a unique solution to \eqref{eq: main estimation optimization} is guaranteed. On the other hand, ${\rm rank}\left(\Psi\right)=\underline{q}$.
% The convergence criterion for the Q-learning is to check on the updates in the estimated vector $s_i$.
% This can be done by the following condition,
In Algorithm \eqref{Alg: Alg1}, instead of getting $q$ samples in each iteration and updating matrix $S$, we update the algorithms using only a single data. Another advantage is that persistent excitation is needed only in the initial iteration.
% Indeed, instead of throwing away the previous data and getting a new set of data, we keep the data and add new data to the previous data.
% Unlike \cite{kiumarsi2017h,al2007model} that use indexes $k$ for time step and $i$ for the algorithm iterations, 
\begin{Remark}
In section \ref{sec: Implement RL Alg}, we only have one index, $i$, since in each iteration we use only a single data. Therefore, we do not require the use of both subscript $i$ and superscript $t$ and only use index $i$.
\end{Remark}

\subsection{Recursive Least Square (RLS)}
Least square (LS) estimation is used when one has an overdetermined system of equations.
% In other words, we have collected more measurements than parameters that we wish to estimate, and, therefore, some of these equations may contradict each other. Thus, we form estimates based on minimizing the mean square (Euclidean norm) error between the measurement vector taken from our data and the measurement vector reproduced from our parameter estimates. However, 
If data is coming in sequentially, we do not have to recompute everything each time a new data point comes in.
% We would still want our estimate to be optimal, too.
Moreover, we can write our new, updated estimate in terms of our old estimate \cite{goel2020recursive}. 

% Consider the model
% \begin{align*}
% {\Psi _{i}}{s_{i +1}} \approx  \Gamma_{i}.
% \end{align*}
% $s_{i+1}$ can be estimated as follows:
% \begin{align*}
% {s}_{i+1} = \arg\min_{s}
% % \sum_{j=-1}^{i+1}
% \left( {\Psi _{i}}s -\Gamma_{i}\right)^T\left( {\Psi _{i}}s -\Gamma_{i}\right).
% \end{align*}
Consider Eq. \eqref{eq: main estimation optimization}. The solution can thus be written as 
% The solution can thus be written as 
\begin{align}
{\Psi _{i}}^T{\Psi _{i}} {s}_{i+1} ={\Psi _{i}}^T\Gamma_{i}.
\label{eq: Sol LS}
\end{align}
By defining ${\Xi_{i}} = \Psi _{i}^T{\Psi _{i}}$, we have
\begin{align}
{\Xi_{i}} = \Psi _{i}^T{\Psi _{i}} =& \Psi _{i-1}^T{\Psi _{i-1}}
% \notag\\&
+{\rm vecv}(\phi_{i}(K_v^i))^T{\rm vecv}(\phi_{i}(K_v^i))
= {\Xi_{i-1}} + {\rm vecv}(\phi_{i}(K_v^i))^T{\rm vecv}(\phi_{i}(K_v^i)).
\label{eq: M inverse}
\end{align}
% and consequently
% \begin{align*}
% % &
% {\Xi_i}{s}_{i+1} =& {\Xi_{i-1}}{ s}_{i+1} + {\rm vecv}(\phi_{i}(K^i))^T{\rm vecv}(\phi_{i}(K^i))s_{i+1}
% \\
%  =& {\Xi_{i-1}}{s}_{i+1} + {\rm vecv}(\phi_{i}(K^i))^T\gamma_{i}\\
%  % = & \Xi_{i-1}{s}_{i}+ {\rm vecv}(\phi_{i}(K^i))^T \\
%  % & \times 
%  % \left({\rm vecv}(\phi_i(K^i))\xi+{\rm vecv}(\phi_{i}(K^i)){s}_{i}\right).
% \end{align*}
Rearranging Eq. \eqref{eq: Sol LS}, we get 
\begin{align*}
{\Xi_i}{s}_{i+1} =& \Psi _{i-1}^T \Gamma_{i-1}
+{\rm vecv}(\phi_{i}(K_v^i))^T\gamma_{i}
= {\Xi_{i-1}}{s}_{i} + {\rm vecv}(\phi_{i}(K_v^i))^T\gamma_{i}.
\end{align*}
% Thus,
% \begin{align*}
% &\Xi_{i} {s}_{i+1} = \Xi_{i-1}{s}_{i}+ {\rm vecv}(\phi_{i}(K^i))^T\left({\rm vecv}(\phi_i(K^i))\xi+{\rm vecv}(\phi_{i}(K^i)){s}_{i}\right).
% \end{align*}
By denoting $M_{i} = \Xi_{i}^{-1}$,
\begin{align*}
{s}_{i+1} = M_{i}\left(\Xi_{i-1}{s}_{i} + {\rm vecv}(\phi_{i}(K_v^i))^T\gamma_{i}\right).
\end{align*}
Plug the above equation into \eqref{eq: M inverse}, it yields
\begin{align*}
{s}_{i+1} =&{s}_{i} - M_{i}\left({\rm vecv}(\phi_{i}(K_v^i))^T{\rm vecv}(\phi_{i}(K_v^i)){s}_{i}
-{\rm vecv}(\phi_{i}(K_v^i))^T\gamma_{i}\right)
\notag\\
=&
{s}_{i} + M_{i}{\rm vecv}(\phi_{i}(K_v^i))^T\left(\gamma_{i}- {\rm vecv}(\phi_{i}(K_v^i)){s}_{i}
\right),
\end{align*}
where $M_{i}$ can be updated in each iteration using Sherman-Morrison formula (\cite{sherman1950adjustment}) as follows:
\begin{align}
M_{i} = M_{i-1} - \frac{M_{i-1}{\rm vecv}(\phi_{i}(K_v^i))^T{\rm vecv}(\phi_{i}(K_v^i))M_{i-1}}{1+{\rm vecv}(\phi_{i}(K_v^i))M_{i-1}{\rm vecv}(\phi_{i}(K_v^i))^T}.
\label{eq: Sherman-Morrison}
\end{align}
The quantity $M_{i}{\rm vecv}(\phi_{i}(K_v^i))^T$ is called the "Kalman Filter Gain", and $\gamma_{i}- {\rm vecv}(\phi_{i}(K_v^i)){s}_{i}$ is called 'innovations' since it compares the difference between a data update and the action given the last estimate. 
% An Implementation Issue: 
% Another concept which is important in the implementation of the RLS algorithm is the computation of $M_{i+1} = Q_{i+1}^{-1}$. 
If the dimension of $\Xi_{i}$ is very large, computation of its inverse can be computationally expensive, so one would like to have a recursion for the $M_{i+1}$ as in \eqref{eq: Sherman-Morrison}.
\begin{Theorem}[Convergence of Algorithm \ref{Alg: Alg1}]
% Given System \eqref{eq: mainSYS}, assume $\left(A,B\right)$ is stabilizable.
% Then, Algorithm \ref{Alg: Alg1} generates a sequence of controls $\{v^j_t, d^j_t ,j = 1,2,3,...\}$ that converges to the optimal feedback controller and disturbance given in \eqref{eq: opt cont gain} and \eqref{eq: opt dist gain} as $j \to \infty$.
Assume that the linear quadratic problem \eqref{eq: mainSYS}-\eqref{eq: value function} is solvable and has a value under the state feedback information structure or equivalently assume there exists a solution to the game's algebraic Riccati recursion \eqref{eq: ricatti recursion}. 
% Algorithm \ref{Alg: Alg1} 
Then, iterating on \eqref{eq: it on S} (equivalent to iterating on \eqref{eq: ricatti recursion}) with
$S^0 = 0$, $K_v^0 =0 $, and $K_d^0 = 0$ converges with $S^i \to S^\star$ and equivalently
$P^i \to P^\star$ where
% .
% converges to the optimal control solution given by \eqref{eq: opt cont gain} and \eqref{eq: opt dist gain} 
the matrix $P^\star$ satisfies the following Recatti equation:
\begin{align}
P&^\star = R_x+{\mathcal A}^TP^\star{\mathcal A} -\begin{bmatrix}
{\mathcal A}^TP^\star{\mathcal B} & {\mathcal A}^TP^\star{\mathcal L}
\end{bmatrix}
\begin{bmatrix}
 R_v+B^TP^\star{\mathcal B} & {\mathcal B}^TP^\star{\mathcal L}\\
 {\mathcal L}^TP^\star{\mathcal B} & {\mathcal L}^TP^\star{\mathcal L}-\gamma^2I
\end{bmatrix}^{-1}
\begin{bmatrix}
{\mathcal B}^TP^\star{\mathcal A} \\ {\mathcal L}^TP^\star{\mathcal A}
\end{bmatrix}.
\label{eq: ricatti equation}
\end{align}
% Eq. \eqref{eq: ricatti recursion}.
\label{Theorem: convegence}
\end{Theorem}
\begin{proof}
 Recall the solution of the problem \eqref{eq: main estimation optimization}:
\begin{align*}
&\Psi _{i}^T{\Psi _{i}}{s_{i + 1}} = \Psi _{i}^T{\Psi _{i}}\xi  + \Psi _{i}^TX^+_{i}p_{i}.
\end{align*}
% By defining
% \begin{align*}
% % \end{align*}
% % \begin{align*}
% {\Xi_{i}} &= \Psi _{i}^T{\Psi _{i}} = \Psi_{i-1}^T{\Psi _{i-1}} 
% +{\rm vecv}(\phi_{i}(K^i))^T{\rm vecv}(\phi_{i}(K^i))\notag\\
% &= {\Xi_{i-1}} + {\rm vecv}(\phi_{i}(K^i))^T{\rm vecv}(\phi_{i}(K^i)),
% \end{align*}
% and ${M_{i}} = {\Xi_{i}^{ - 1}}$, 
The following equation can be concluded:
\begin{equation}
{s_{i + 1}} = \xi  + {M_{i}}{\Omega_{i}}{p_{i}},
\label{eq: s estimation equation}
\end{equation}
where
\begin{align*}
    {\Omega _{i}} &= \Psi _{i}^TX^+_{i} = \Psi _{i-1}^TX^+_{i-1} + {\rm vecv}(\phi_{i}(K_v^i))^T{\rm vecv}(x_{i+1})
    % \notag\\&
= {\Omega_{i-1}} + {\rm vecv}(\phi_{i}(K_v^i))^T{\rm vecv}(x_{i+1}).
\end{align*}
Using Sherman–Morrison formula \eqref{eq: Sherman-Morrison} and
% , we can expand $M_{i}$ as follow
% \begin{align}
%   {M_{i}} =
%   {M_{i-1}} - \frac{{{M_{i-1}}{\rm vecv}(\phi_{i}(K^i))^T{\rm vecv}(\phi_{i}(K^i)){M_{i-1}}}}{{1 + {\rm vecv}(\phi_{i}(K^i)){M_{i-1}}{\rm vecv}(\phi_{i}(K^i))^T}}.
%   \label{eq: M_Sherman}
% \end{align}
defining ${W_{i}} = {M_{i}}{\Omega _{i}}$, we have
\begin{align*}
{W_{i}}&= {M_{i}}{\Omega _{i}}
% \notag \\&
= \left(    {M_{i-1}} - \frac{{{M_{i-1}}{\rm vecv}(\phi_{i}(K_v^i))^T{\rm vecv}(\phi_{i}(K_v^i)){M_{i-1}}}}{{1 + {\rm vecv}(\phi_{i}(K_v^i)){M_{i-1}}{\rm vecv}(\phi_{i}(K_v^i))^T}} \right)
% \notag\\&\times
\left({\Omega _{i-1}} + {\rm vecv}(\phi_{i}(K_v^i))^T{\rm vecv}(x_{i+1})\right)\notag\\
&= {W_{i-1}} - \frac{{{M_{i-1}}{\rm vecv}(\phi_{i}(K_v^i))^T{\rm vecv}(\phi_{i}(K_v^i))}}{{1 + {\rm vecv}(\phi_{i}(K_v^i)){M_{i-1}}{\rm vecv}(\phi_{i}(K_v^i))^T}}{W_{i-1}}
% \notag\\ &
+ M_{i}{\rm vecv}(\phi_{i+1}(K_v^i))^T{\rm vecv}(x_{i+1})\notag \\
 &= {W_{i-1}} - \frac{{{M_{i-1}}{\rm vecv}(\phi_{i}(K_v^i))^T{\rm vecv}(\phi_{i}(K_v^i))}}{{1 + {\rm vecv}(\phi_{i}(K_v^i)){M_{i-1}}{\rm vecv}(\phi_{i}(K_v^i))^T}}{W_{i-1}} 
 % \notag\\&
 +\frac{{{M_{i-1}}{\rm vecv}(\phi_{i}(K_v^i))^T}}{{1 + {\rm vecv}(\phi_{i}(K_v^i)){M_{i-1}}{\rm vecv}(\phi_{i}(K_v^i))^T}}
 {\rm vecv}(x_{i+1}).
%  \label{eq: W_recursion}
\end{align*}
Recall $x_{i+1} = {\mathcal A}x_i +{\mathcal B}K_v^i x_i + {\mathcal L}d_i$. Hence, ${\rm vecv}(x_{i+1}) ={\rm vecv}({\mathcal A}x_i +{\mathcal B}K_v^i x_i + {\mathcal L}d_i)$. Note that ${\rm vecv}(x_{i+1})$ can be partitioned as ${\rm vecv}(\phi_{i}(K_v^i)) 
f
({\mathcal A},{\mathcal B},{\mathcal L})$,
% \begin{align*}
% {\rm vecv}(x_{i+1}) =& {\rm vecv}(({\mathcal A}+{\mathcal B}K_c^i+{\mathcal L}K_d^i)x_i) \notag\\ =& {\rm vecv}(\phi_{i}(K^i)) f({\mathcal A},{\mathcal B},{\mathcal L}),
% \end{align*}
where $f({\mathcal A},{\mathcal B},{\mathcal L})$ is a matrix that its entities are a function of the entities of the matrices ${\mathcal A}$, ${\mathcal B}$, and ${\mathcal L}$. Therefore, $W_i$ can be written as follows: 
\begin{align*}
{W_{i}} 
% &{W_{i-1}} - \frac{{{M_{i-1}}{\rm vecv}(\phi_{i}(K^i))^T{\rm vecv}(\phi_{i}(K^i))}}{{1 + {\rm vecv}(\phi_{i}(K^i)){M_{i-1}}{\rm vecv}(\phi_{i}(K^i))^T}}{W_{i-1}} \notag\\
%  &+\frac{{{M_{i-1}}{\rm vecv}(\phi_{i}(K^i))^T}}{{1 + {\rm vecv}(\phi_{i}(K^i)){M_{i-1}}{\rm vecv}(\phi_{i}(K^i))^T}}
%  {\rm vecv}(x_{i+1}) \notag\\
 = &{W_{i-1}} - \frac{{{M_{i-1}}{\rm vecv}(\phi_{i}(K_v^i))^T{\rm vecv}(\phi_{i}(K_v^i))}}{{1 + {\rm vecv}(\phi_{i}(K_v^i)){M_{i-1}}{\rm vecv}(\phi_{i}(K_v^i))^T}}{W_{i-1}}
 % \notag\\
 % &
 +\frac{{{M_{i-1}}{\rm vecv}(\phi_{i}(K_v^i))^T{\rm vecv}(\phi_{i}(K_v^i))f({\mathcal A},{\mathcal B},{\mathcal L})}}{{1 + {\rm vecv}(\phi_{i}(K_v^i)){M_{i-1}}{\rm vecv}(\phi_{i}(K_v^i))^T}}.
\end{align*}
By subtracting $f({\mathcal A},{\mathcal B},{\mathcal L})$ from both sides of above equation, we have:
\begin{align*}
&{W_{i}} - f({\mathcal A},{\mathcal B},{\mathcal L})={W_{i-1}}-f({\mathcal A},{\mathcal B},{\mathcal L})
% \\&
-\frac{{{M_{i-1}}{\rm vecv}(\phi_{i}(K_v^i))^T{\rm vecv}(\phi_{i}(K_v^i))}}{{1 + {\rm vecv}(\phi_{i}(K_v^i)){M_{i-1}}{\rm vecv}(\phi_{i}(K_v^i))^T}}
% \\&\times
\left ({W_{i-1}} -f({\mathcal A},{\mathcal B},{\mathcal L})\right).
\end{align*}
Let's denote ${\hat W}_{i-1} = W_{i-1}- f({\mathcal A},{\mathcal B},{\mathcal L})$. Above equation yields
\begin{align*}
{\hat W}_{i}
% &={\hat W}_{i-1}-\frac{{{M_{i-1}}{\rm vecv}(\phi_{i}(K^i))^T{\rm vecv}(\phi_{i}(K^i))}}{{1 + {\rm vecv}(\phi_{i}(K^i)){M_{i-1}}{\rm vecv}(\phi_{i}(K^i))^T}}
% {\hat W}_{i-1}\\
& = \left( I-\frac{{{M_{i-1}}{\rm vecv}(\phi_{i}(K_v^i))^T{\rm vecv}(\phi_{i}(K_v^i))}}{{1 + {\rm vecv}(\phi_{i}(K_v^i)){M_{i-1}}{\rm vecv}(\phi_{i}(K_v^i))^T}}\right){\hat W}_{i-1}
% \\&
=M_{i}M_{i-1}^{-1}{\hat W}_{i-1}.
% = M_{i}M_{0}^{-1}{\hat W}_{0}.
\end{align*}
Considering ${\hat W}_1 = M_1M_0^{-1}{\hat W}_0$, 
we can have ${\hat W}_2$ as follows: 
\begin{align*}
{\hat W}_2 = M_2M_1^{-1}{\hat W}_1 = M_2M_1^{-1}M_1M_0^{-1}{\hat W}_0= M_2M_0^{-1}{\hat W}_0.
\end{align*}
As a result, we can conclude
${\hat W}_{i}=M_{i}M_{i-1}^{-1}{\hat W}_{i-1} = M_{i}M_{0}^{-1}{\hat W}_{0}$.
% \begin{align*}
% {\hat W}_{i}&=M_{i}M_{i-1}^{-1}{\hat W}_{i-1}\\
% {\hat W}_1 &= M_1M_0^{-1}{\hat W}_0\\
% {\hat W}_2 &= M_2M_1^{-1}{\hat W}_1 = M_2M_0^{-1}{\hat W}_0\\
% \vdots\\
% {\hat W}_{i}&=M_{i}M_{0}^{-1}{\hat W}_{0}
% \end{align*}
% In Lemma \ref{lemma1}, we will investigate the convergence $W$.
% \begin{Lemma}
% \label{lemma1}
% Given any $\epsilon >0$ there exist an $N_0$ such that for all $N > N_0$, we have $|| W_{N}-W_{N-1}||_2<\epsilon$.
% \end{Lemma}
% \begin{proof}
% \begin{align*}
% {\hat W}_{i}&=M_{i}M_{i-1}^{-1}{\hat W}_{i-1}\\
% {\hat W}_1 &= M_1M_0^{-1}{\hat W}_0\\
% {\hat W}_2 &= M_2M_1^{-1}{\hat W}_1 = M_2M_0^{-1}{\hat W}_0\\
% \vdots\\
% {\hat W}_{i}&=M_{i}M_{0}^{-1}{\hat W}_{0}
% \end{align*}
By expanding ${\hat W}_i$, we have:
\begin{align*}
W_{i} = & M_{i}M_{0}^{-1} \left(W_{0} - f({\mathcal A},{\mathcal B},{\mathcal L})\right) + f({\mathcal A},{\mathcal B},{\mathcal L})\\
= & M_{i}\Psi_0^T\Psi_0\left (M_0\Psi_0^TX^+_0 - f({\mathcal A},{\mathcal B},{\mathcal L}) \right) +f({\mathcal A},{\mathcal B},{\mathcal L}) \\
  = & M_{i}\Psi_0^T\Psi_0\left ((\Psi_0^T\Psi_0)^{-1}\Psi_0^TX^+_0 -f({\mathcal A},{\mathcal B},{\mathcal L}) \right) 
  % \\ &
  +f({\mathcal A},{\mathcal B},{\mathcal L}) \\
   = & M_{i}(\Psi_0^TX^+_0 -\Psi_0^T\Psi_0f({\mathcal A},{\mathcal B},{\mathcal L})) + f({\mathcal A},{\mathcal B},{\mathcal L})\\
  = & M_{i}\Psi_0^T(X^+_0 -\Psi_0f({\mathcal A},{\mathcal B},{\mathcal L})) + f({\mathcal A},{\mathcal B},{\mathcal L}).
\end{align*}
The terms $X_0^+$ and $\Psi_0f({\mathcal A},{\mathcal B},{\mathcal L})$ are vectorized form of $x_{t+1}$ and  ${\mathcal A}{x}_t+{\mathcal B}{v}_t+\mathcal{L} {d}_t$, respectively, for the initial $q$ time steps, respectively. Hence,
$$X^+_0 -\Psi_0f({\mathcal A},{\mathcal B},{\mathcal L}) = 0.$$ Consequently, it results in $W_{i} = f({\mathcal A},{\mathcal B},{\mathcal L})$.
% \end{proof}
Therefore, by reconstructing \eqref{eq: s estimation equation}, it follows \eqref{eq: it on S} and consequently 
the following Ricatti recursion:
\begin{align}
P&^{i+1} = R_x+{\mathcal A}^TP^i{\mathcal A} -
% \notag\\&
\begin{bmatrix}
{\mathcal A}^TP^i{\mathcal B} & {\mathcal A}^TP^i{\mathcal L}
\end{bmatrix}
\begin{bmatrix}
 R_v+B^TP^i{\mathcal B} & {\mathcal B}^TP^i{\mathcal L}\\
 {\mathcal L}^TP^i{\mathcal B} & {\mathcal L}^TP^i{\mathcal L}-\gamma^2I
\end{bmatrix}^{-1}
% \notag\\ &\times
\begin{bmatrix}
{\mathcal B}^TP^i{\mathcal A} \\ {\mathcal L}^TP^i{\mathcal A}
\end{bmatrix}.
\end{align}
By using {\it Lemma 4.1 and Theorem 4.2} in \cite{stoorvogel1994discrete}, it is shown that iterating on \eqref{eq: ricatti recursion} with $P_0=0$ converges to $P^*$. 
\end{proof}
\subsection{Computational Complexity Analysis}
Recall $\underline{q} = {{(m_1 + m_2+m_3)(m_1 + m_2+m_3+ 1)}/ 2}$ as the number of parameters to be estimated. In both classical Q-learning and the proposed algorithm, the number of parameters being estimated is similar. 
% Computational complexity in each iteration of Algorithms \ref{Alg: Alg1} and classical Q-learning in the literature: ${\mathcal D}$: the number of parameters to be estimated, $m_1$, $m_2$, and $m_3$: the dimensions of the state, control input, and disturbance input, $q$ : The exploration length in the data collection.
For the sake of comparison, assume $q =\underline{q}$. for the initial iteration both of the algorithms have a computational complexity of order ${\mathcal O}(\underline{q}^3)$ while in the rest of the iterations, Algorithm \ref{Alg: Alg1} has a computational complexity of order ${\mathcal O}(\underline{q}^2)$, unlike classical Q-learning that has ${\mathcal O}(\underline{q}^3)$ order of computational complexity. In \cite{al2007model,kiumarsi2017h}), to update the parameters of the critic network, at least $\underline{q}$ data is required, and because there is a batch of data in each iteration, a pseudo-inverse (with the computational complexity of $\mathcal{O}(\underline{q}^3)$) in each iteration must be computed. In contrast, we emphasize sample complexity and use only a {\it single data} to update the parameters of the critic network. It is a huge advantage for systems that have long time steps or when acquiring data is not trivial. On the order of computational complexity, using the key equation
$${s}_{i+1} ={s}_{i} + M_{i}{\rm vecv}(\phi_{i}(K_v^i))^T\left(\gamma_{i}- {\rm vecv}(\phi_{i}(K_v^i)){s}_{i}
\right),
$$
the computational complexity of $M_{{i}_{\underline{q}\times \underline{q}}}{\rm vecv}(\phi_{i}(K_v^i))^T_{{\underline{q}\times 1}}$ is $\mathcal{O}(\underline{q}^2)$ (considering $\gamma_{i}- {\rm vecv}(\phi_{i}(K_v^i)){s}_{i}$ is a scalar).
%  is $\mathcal{O}(\underline{q})$ (a vector times a scalar), 
The computational complexity of the key equation reduces to the computational complexity of calculating $M_i$ in \eqref{eq: Sherman-Morrison}. The computational complexity of calculating the column vector $M_{{i-1}_{\underline{q}\times \underline{q}}}{\rm vecv}(\phi_{i}(K_v^i))^T_{{\underline{q}\times 1}}$ and the row vector ${\rm vecv}(\phi_{i}(K_v^i))_{{1 \times \underline{q}}}M_{{i-1}_{\underline{q}\times \underline{q}}}$are $\mathcal{O}(\underline{q}^2)$. Considering the computational complexity of the scalar ${\rm vecv}(\phi_{i}(K_v^i))_{1 \times \underline{q}}M_{{i-1}_{\underline{q}\times \underline{q}}}{\rm vecv}(\phi_{i}(K_v^i))^T_{\underline{q}\times 1}$ is $\mathcal{O}(\underline{q}^2)$, therefore, the computational complexity of calculating $M_{i}$ is $\mathcal{O}(\underline{q}^2)$, and consequently, the computational complexity of calculating $s_{i+1}$ is $\mathcal{O}(\underline{q}^2)$.

\section{Autonomous Mobility-on-Demand (AMoD) Model}
% \begin{Proposition} 
% \end{Proposition}
% \begin{proof}
% \end{proof}
\label{sec: AMoD system modeling}
% This section explains how a directed graph can be transformed into a complete graph under a particular assumption. Also,
In this section, a discrete-time linear dynamic model is formulated for the AMoD system.
% At the end of this section, some properties of the proposed model are discussed.
% \subsection{Model}
We relax the model in \cite{carron2019scalable} by considering origin-destination demand. 
The linear discrete-time time-delay dynamic system is as follows:
\begin{subequations}
  \begin{align}
  \label{Apr_LinearModel_a}
w^{rs}\left ( t+1 \right )= & w^{rs}\left ( t \right ) +d^{rs}\left ( t \right )-{U}^{rs}\left ( t \right )
% ,{\,\forall  r , s \in N} 
\\
  \label{Apr_LinearModel_b}
p_{r}\left ( t+1 \right )=&  p_{r}\left ( t\right )-\sum_{s\in N} {\left( {U}^{rs}\left ( t \right ) 
+{R}^{rs}\left ( t \right )\right ) }+\sum_{q\in N} {\left(\frac{{g}^{qr}\left ( t\right )}{T_{qr}} 
\right )}
% ,{\,\forall \, r , s \in N}
\\\label{Apr_LinearModel_c}
g^{rs}\left ( t+1 \right )=&\left ( 1- \frac{1}{T_{rs}}\right )g^{rs}\left ( t\right )+ {U}^{rs}\left ( t \right )+{R}^{rs}\left ( t \right ),
\end{align}
\label{Apr_LinearModel}
\end{subequations}
for ${\forall \, r , s \in N}$ where state variable $w^{rs}$ denotes the waiting customers at $r$ aiming to go to $s$.
% , i.e., customers willing to travel from station $r$ to station $s$.
State variable $p_r$ characterizes the waiting or available vehicles at station $r$.
State variable $g^{rs}$ denotes vehicles moving along the link $\left\{ r,s \right\}$, including both customer-carrying and rebalancing vehicles. Control input $U^{rs}$ is the number of available vehicles at station $r$ with a customer that will be dispatched to link $\left\{ r,s \right\}$. $R^{rs}$ is the number of available vehicles at station $r$ that will be dispatched to link $\left\{ r,s \right\}$ for rebalancing. The term $d^{rs}(t)$ represents the arrival of customers in a time step given by the realization of a Poisson process of parameter $\lambda^{rs}$.
% Equation \eqref{Apr_LinearModel_a} describes the evolution of the passengers at $r$ headed to $s$, equation \eqref{Apr_LinearModel_b} captures the evolution of the available vehicle, and equation \eqref{Apr_LinearModel_c} describes the evolution of link flow dynamics for customer-carrying and rebalancing vehicles.
Note that each vehicle serves only one customer request at a time, i.e., sharing/pooling is not considered.
We also assume that the travel times $T_{rs}$ are constant and exogenous. The reason is that the number of AMoD vehicles is much less than the rest of the traffic.
Model \eqref{Apr_LinearModel} is derived using a first-order lag approximation of the time delays. It is assumed that the number of vehicles exiting a link is proportional to the number of vehicles on that link. In other word, at each time instant $t$, the quantity $g^{rs}\left ( t \right )/T_{rs}$ leaves the link $\left \{ r,s \right \}$. Therefore, 
${U}^{rs}\left ( t-T_{rs} \right ) +{R}^{rs}\left ( t - T_{rs}\right )$ can be replaced by $g^{rs}\left ( t \right )/T_{rs}$.

This AMoD system is subject to some constraints that enforce the non-negativity of state and control input variables. The global system associated with graph $G$ is represented as
  \begin{equation}
  {x}_{t+1}={\mathcal A}{x_{t}} + {\mathcal B}{v_t}+\mathcal{L} {{d_t}},
       \label{SystemVector}
   \end{equation}
where the vector of all state variables ${x_t}\in {\mathbb{R}^{2n^2-n}}$ is $\begin{bmatrix}{w}\left ( t\right )^T,{p}\left ( t\right )^T
,{g}\left ( t\right )^T\end{bmatrix}^T$ and the vector of all control input variables ${v_t} \in \mathbb{R}^{{2n(n-1)}}$ is defined as ${v_t}=\begin{bmatrix}
{U}\left( t\right )^T,{R}\left ( t\right )^T
\end{bmatrix}^T$. ${{d_t}} \in {\mathbb{R}^{n(n-1)}}$ represents arriving customers. Matrices ${\mathcal A}$, ${\mathcal B}$, and ${\mathcal L}$ can be written as below:
\begin{align}
&\mathcal A= 
\begin{bmatrix}
 I_{n(n-1)} & 0 & 0\\ 
 0 & I_n & E_{\rm in}\tilde{T}^{-1}\\ 
 0 & 0 & I_{n(n-1)}-\tilde{T}^{-1}
\end{bmatrix}, \quad \mathcal B=
\begin{bmatrix}
-I_{n(n-1)} & 0\\ 
-E_{\rm out}& -E_{\rm out}\\ 
I_{n(n-1)} & I_{n(n-1)}
\end{bmatrix}, \quad \mathcal{L} = \begin{bmatrix}
I_{n(n-1)}\\ 
0\\
0
\end{bmatrix}.
 \end{align}
If graph $G$ is strongly
connected and $d^{rs}$ = $\lambda^{rs}$ for $\forall \left \{r, s\right \} \in {A}$, where $\lambda^{rs}$ represents the Poisson arrival rate for the link $\left \{r, s\right \}$, then equilibrium points of system \eqref{SystemVector} are given by 
${\bar{ x}}=\left ({\bar{ w}},{\bar{p}}
,{\bar{ g}} \right )$, where ${\bar{w}}$ and ${\bar{ p}}$
can be any arbitrary positive vector, ${\bar{ g}}=\tilde{T} \left (  {{ \lambda}} + {\bar{ R}}\right )$, ${\bar{ U}}={{\lambda}}$, and ${\bar{ R}}$ satisfies 
$E\left ({\bar{R}}+{{ \lambda}}\right )=0$.
If the number of nodes, $n$, is greater than 2, there will be an infinite number of equilibrium points. Also, the desired equilibrium point that minimizes the number of rebalancing, ${\bar R}^{\star}$, can be found by solving 
% an
the following
optimization problem:
% \end{Corollary}
\begin{subequations}
\begin{flalign}
\label{rebalnce_optimization_cost}
\mathop {\min {\mkern 1mu} }\limits_{\bar R}\quad& \left\| {\tilde T }^{\frac{1}{2}}{\bar R} \right\|_2^2\\
\rm{s.t.}\quad&E\left ({{\bar R}}+{{\lambda}}\right ) = 0,\quad{{\bar R}}\geq 0.
\end{flalign}
\label{OptForRebalance}
\end{subequations}
By changing the coordinates of \eqref{SystemVector}, we aim to regulate the AMoD system around the desired equilibrium points.

\section{Simulation Study}
\label{sec: results}
\begin{comment}
\begin{figure*}[thbp]
  \centering
  \begin{subfigure}{0.4\textwidth}
    \centering
  \includegraphics[width=1.6\textwidth]{Fig/Layout-N.pdf}
  \caption{ }
  \label{fig:Campus Network}
  \end{subfigure}
\hfill
\begin{subfigure}{0.35\textwidth}
    \centering
  \includegraphics[width=1\textwidth]{Fig/IEEE-T-ITS-MU.pdf}
 \caption{ }
   \label{fig:Campus zones}
  \end{subfigure} 
  	\caption{(a) The University of Minnesota-Twin Cities campus network map.% for testing the proposed model and control strategy.
  	%Zone 1 denotes the west bank of the Minneapolis campus, Zones 2, 3, 4, and 5 represent the east bank of the Minneapolis campus, and Zone 6 denotes the St. Paul campus.
  	Zones (shown with different symbols) have been decided using the k-mean algorithm and travel data; (b) The University of Minnesota zones.}
  	 % \label{fig:WNetwork}
  	 \end{figure*}
\end{comment}
We first introduce a network for the test we perform. 
% Second, a sensitivity analysis is carried out to tune the control horizon and the dispatching and rebalancing time interval. Then, we present another state-of-the-art rebalancing algorithm to be compared with the MPC framework.
Then, we apply Algorithm \ref{Alg: Alg1} developed in Section \ref{sec: Implement RL Alg} to obtain optimal control, disturbance actions, and the value function parameters in time.
\subsection{Studied Network}
% The University of Minnesota-Twin Cities (UMN) 
% A real network
% shown in Fig. \ref{fig:Campus Network}
% is considered as a case study on which to perform tests.
% There are 132 spots (buildings) demanded by the customers.
% It is

\begin{figure}[thpb]
  \centering
  \includegraphics[width=0.6
\textwidth]{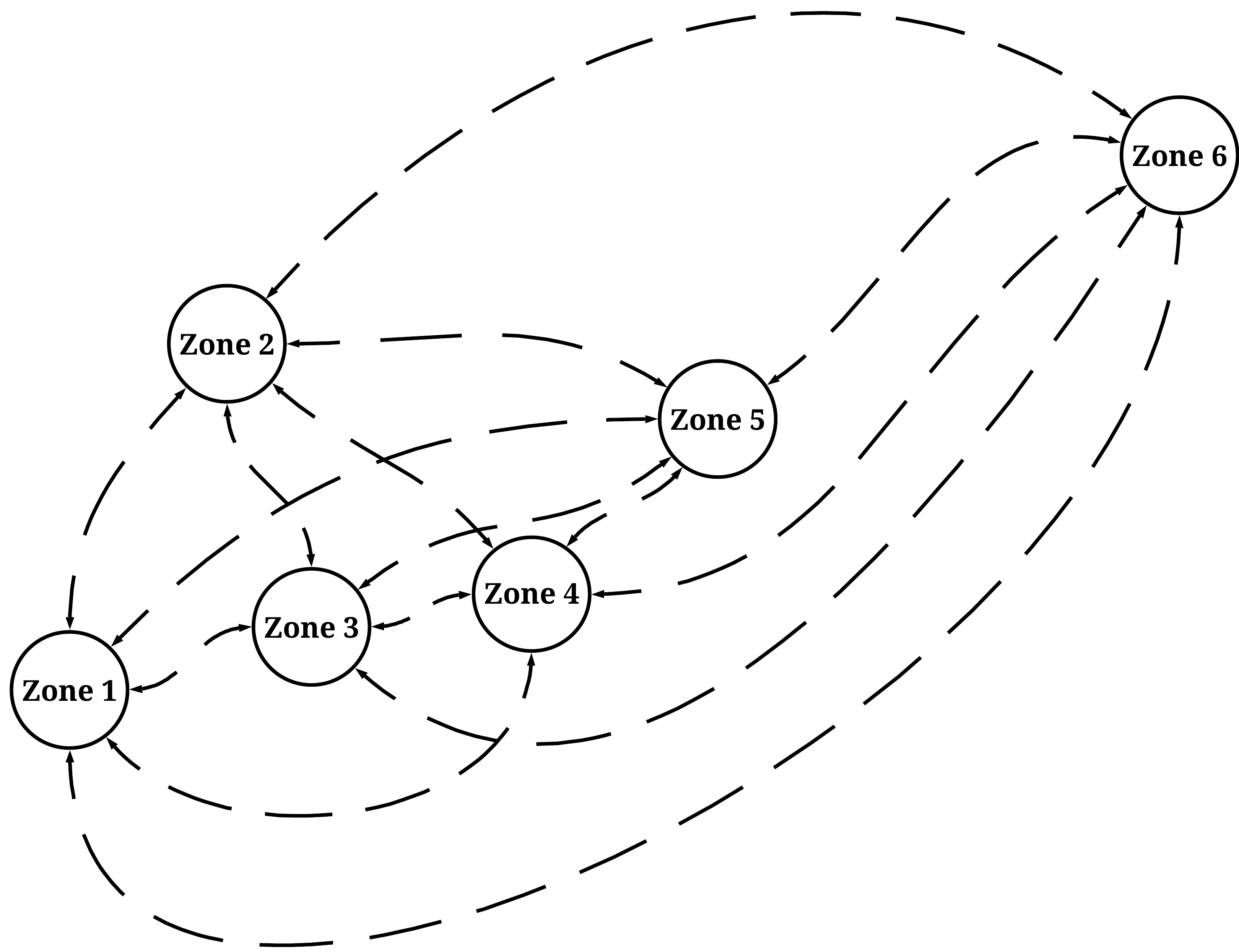}
%   	\caption{The University of Minnesota zones.}
  	\caption{The network zones.}

  \label{fig:Campus zones}
\end{figure}

The University of Minnesota-Twin Cities (UMN) campus
network is considered as the site on which to perform the test. The network we consider is partitioned into six zones.
% using the k-mean algorithm based on the origin-destination demands extracted from trip demand data. AMoD is considered the unique provider of mobility services.
% As a result of the partitioning, we produce a digraph with $n = 6$ vertices and $m = 30$ links; the graph vertices superimposed on the map are shown in Fig. \ref{fig:Campus zones}.
A digraph with $n = 6$ vertices and $m = 30$ links is produced by partitioning; the graph vertices are superimposed on the map shown in Fig. \ref{fig:Campus zones}. It should be noted that the rebalancing
performance is certainly affected by partitioning, but a detailed analysis is beyond the scope of this article.
% \begin{figure}[h]
%   \centering
%  \includegraphics[width=.49
%  \textwidth]{Fig/IEEE-T-ITS-MU.pdf}
%   \caption{The University of Minnesota zones.}
%   \label{fig:Campus zones}
% \end{figure}
% Figure \ref{fig:demand_histogram} shows the histogram of the daily demand for the UMN’s campuses. The peak occurs between 11:00 AM and 1:00 PM.
% \begin{figure}[h]
%   \centering
%  \includegraphics[width=0.49\textwidth]{Fig/Travel_demand_time.pdf}
%   \caption{Histogram of the daily demand for the UMN's campuses.}
%   \label{fig:demand_histogram}
% \end{figure}

% \subsection{Adaptive Real-time Rebalancing}
% To further investigate the performance of the MPC controller, a comparison is carried out with a state-of-the-art rebalancing algorithm, namely adaptive real-time rebalancing (ARR), \cite{pavone2012robotic}. In this approach, the optimal rebalancing policy can be found as the solution to a linear program. This algorithm minimizes the average number of empty (rebalancing) vehicles traveling in the network. 
\begin{figure}[thpb]
  \centering
 \includegraphics[width=.6\textwidth]{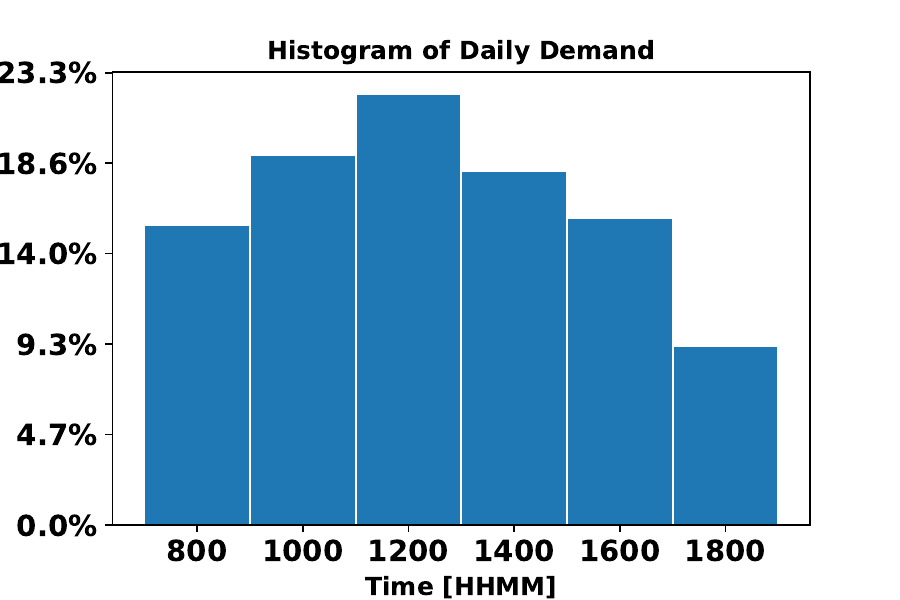}
  \caption{Histogram of the daily demand for the UMN's campuses.}
  \label{fig:demand_histogram}
\end{figure}

Figure \ref{fig:demand_histogram} shows the histogram of the daily demand for UMN’s campuses. The peak occurs between 11:00 AM and 1:00 PM.
Since the demand represents the intra-zonal trips on the campuses (not commutes to the campus), it does not necessarily follow the typical morning and afternoon peaks.

\subsection{Case Study}
% Two case studies are considered to evaluate the performance of the proposed algorithm. Comparisons are performed by applying ... and MPC. In Case study 1, the horizon $N_{\rm {hor}}$ in the Problem \eqref{Main_optimization} is 360, and we consider each time step equal to two minutes, while in Case study 2, the horizon $N_{\rm {hor}}$ is 240, and we consider each time step equal to three minutes. It means dispatching customers and rebalancing are performed every two minutes and three minutes in Case studies 1 and 2. We choose control horizon $N_{\rm c} = 7$ for both case studies. 
% The simulation time for Case studies 1 and 2 are about 10 and 6 minutes, respectively.  

A 12-hour historical trip dataset
% from the UMN 
is considered for the case study. We consider each time step equal to two minutes. So, the number of iterations is 360.
Some origin-destination pairs are used more frequently than others, which implies a significant imbalance in demand.
% The total number of vehicles available is 400.
% % , which serves 15,800 mobility requests.
% The reference being tracked with the MPC controller is recomputed via Problem \eqref{OptForRebalance} every 2 hours implementing a quadratic cost. 
% Afterward, the MPC algorithm is tested with two different scenarios: MPC-QP and MPC-LP. MPC-QP denotes MPC if both costs \eqref{maincost} and \eqref{rebalnce_optimization_cost} are quadratic, while MPC-QP represents MPC with a linear cost in \eqref{maincost} and \eqref{rebalnce_optimization_cost}.
% In Proposition \ref{well-posedness}, we showed that the 
The number of vehicles is constant at each time step (including equilibrium) and is equal to ${\bf 1}_{n}^{T}{p}\left ( t \right ) +{\bf 1}_{n(n-1)}^{T}{ g}\left ( t \right )$  (\cite{pavone2012robotic,carron2019scalable}). Therefore, $\underline{M}= {\bf 1}_{n(n-1)}^{T}{\bar{g}}={T^T} \left (  {{\lambda}} + {\bar{R}}\right )$ can be considered a lower band for the fleet size.
The origin and destination of every trip in the travel data are subsequently assigned to the corresponding zones in the graph. We used Dijkstra’s algorithm \cite{dijkstra1959note} to compute the shortest path between the zone centers on a real road network (Google Maps).
Initial conditions for the AMoD model are ${x}_0=\begin{bmatrix}
0^T_{\frac{n(n-1)}{2}} & \frac{\underline{M}}{n}{1^T_{n}} & 0^T_{\frac{n(n-1)}{2}}  
\end{bmatrix}^T$.
% where ${\bf p}(0) = \frac{\underline{M}}{n} \begin{bmatrix}
% 1^T_{n}  
% \end{bmatrix}^T$.
% Note that, in the simulation, $1.2\underline{M}$ is considered as the fleet size.
% It is important to notice that all the rebalancing algorithms considered need a lower level matching policy, allocating the open requests to individual vehicles. For ARR, the implemented matching policy is global, whereas for both MPC algorithms, it is local, i.e., the matching happens within the same station. 
% Note that no congestion effects are considered in the simulation tests. 
In the performed simulation, no congestion
effects have been considered, i.e., travel times are considered exogenous. If congestion is considered in the model, travel times are endogenous and a function of the policies performed by Algorithm \ref{Alg: Alg1}. In that case, the model is no longer linear and a detailed analysis is beyond the scope of this article.
% The algorithms are compared using the following metrics: the average queue length, average waiting time, i.e., the average time a customer has to wait before being served; and the average empty distance, i.e., the number of miles run by cars without customers.
The average queue length, the average number of rebalancing vehicles, and the average number of customer-carrying vehicles are the metrics that we are interested in investigating using Algorithm \ref{Alg: Alg1}. 
% The eigenvalues of $\mathcal{A}$ are ${\rm Spec}(\mathcal{A}) = {\rm Spec}(I_{n(n-1)}) \cup  {\rm Spec}(I_{n}) \cup
% {\rm Spec}(I_{n(n-1)}-\tilde{T}^{-1})$.
The disturbance attenuation $\gamma$ is selected to be 0.1. Let $W_{\lambda} = 0.01I$. Weights matrices $R_x$ and $R_v$ are chosen as 
% follow:
% \begin{equation}
$
 R_x= \begin{bmatrix}
{\tilde {\bf\lambda}} & 0 & 0\\ 
0 & 0 & 0\\ 
0 & 0 & 0
\end{bmatrix}
$ and $
R_v =\begin{bmatrix}
{\rho\tilde T} & 0 \\ 
0 & {\rho\tilde T}
\end{bmatrix},$
where $\rho = 0.05$. The reference being tracked ($ \lambda$, ${\bar R}^{\star}$) is recomputed via Problem \eqref{OptForRebalance} every 2 hours (60 iterations). Therefore, $R_x$ will be changing every 60 iterations. The recursive least-squares algorithm is used to tune the parameters of the critic network online. The parameters of the actions networks are updated according to \eqref{eq: opt cont gain} and \eqref{eq: opt dist gain}.
% \begin{figure}[thpb]
%   \centering
%  \includegraphics[width=.48
%  \textwidth]{Fig/norm_KK_d_error.pdf}
%   \caption{Convergence of the disturbance action network parameters ($K_d$).}
%   \label{fig: Disterbance Gain}
% \end{figure}
\begin{figure}[thbp]
%   \centering
% \hfill
\begin{subfigure}{0.5\textwidth}
    \centering
  \includegraphics[width= 0.97\textwidth]{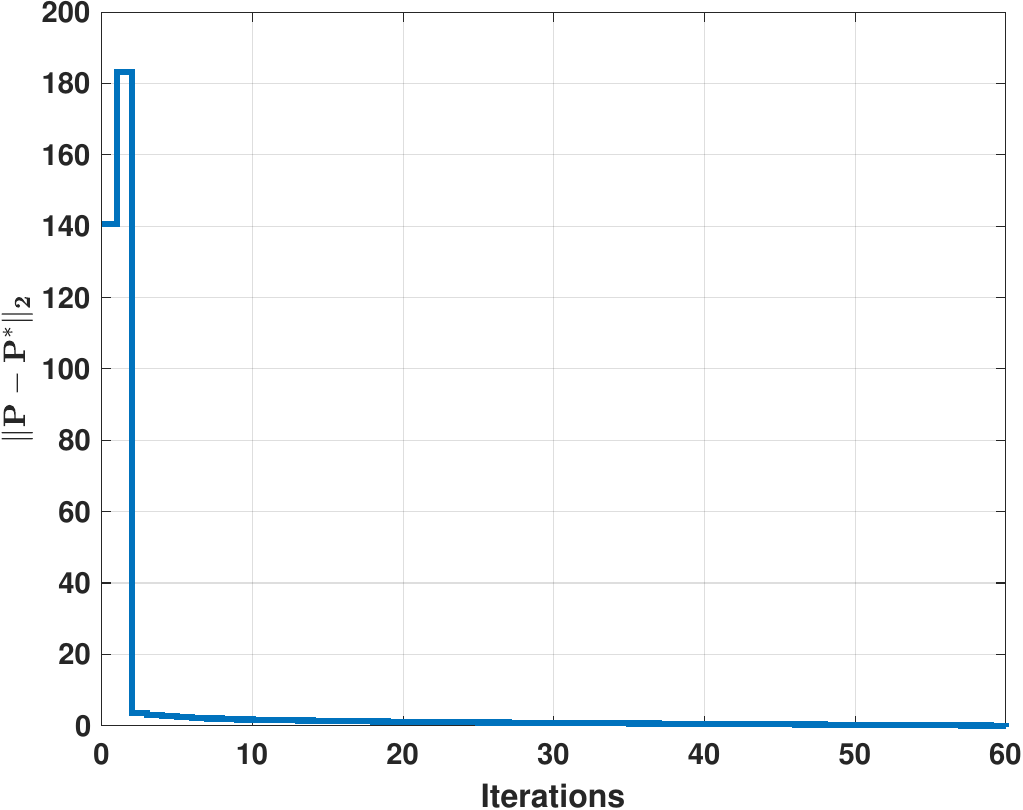}
 \caption{Online convergence of $P$.}
  \label{fig: norm P 0_60}  \end{subfigure}
\hfill
  \centering
\begin{subfigure}{0.5\textwidth}
    \centering
  \includegraphics[width=.97\textwidth]{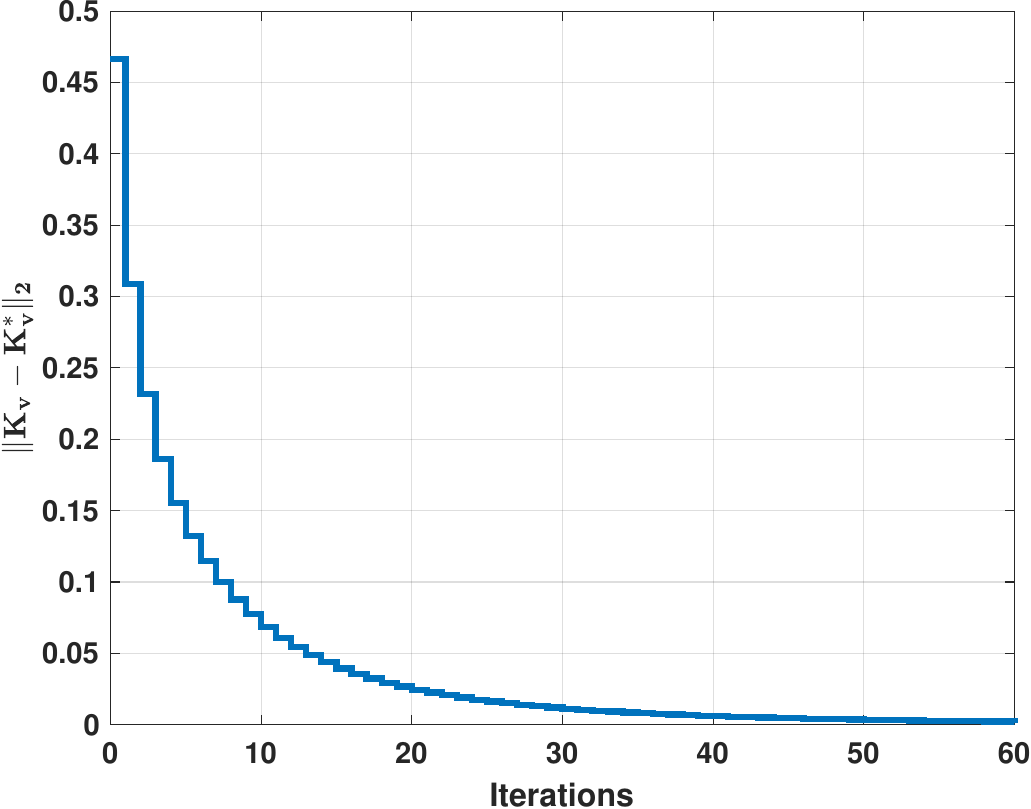}
 \caption{Convergence of the control action network parameters ($K_v$).}
  \label{fig: norm Kc 0_60}
  \end{subfigure}
\hfill
\begin{subfigure}{0.5\textwidth}
    \centering
  \includegraphics[width=.97\textwidth]{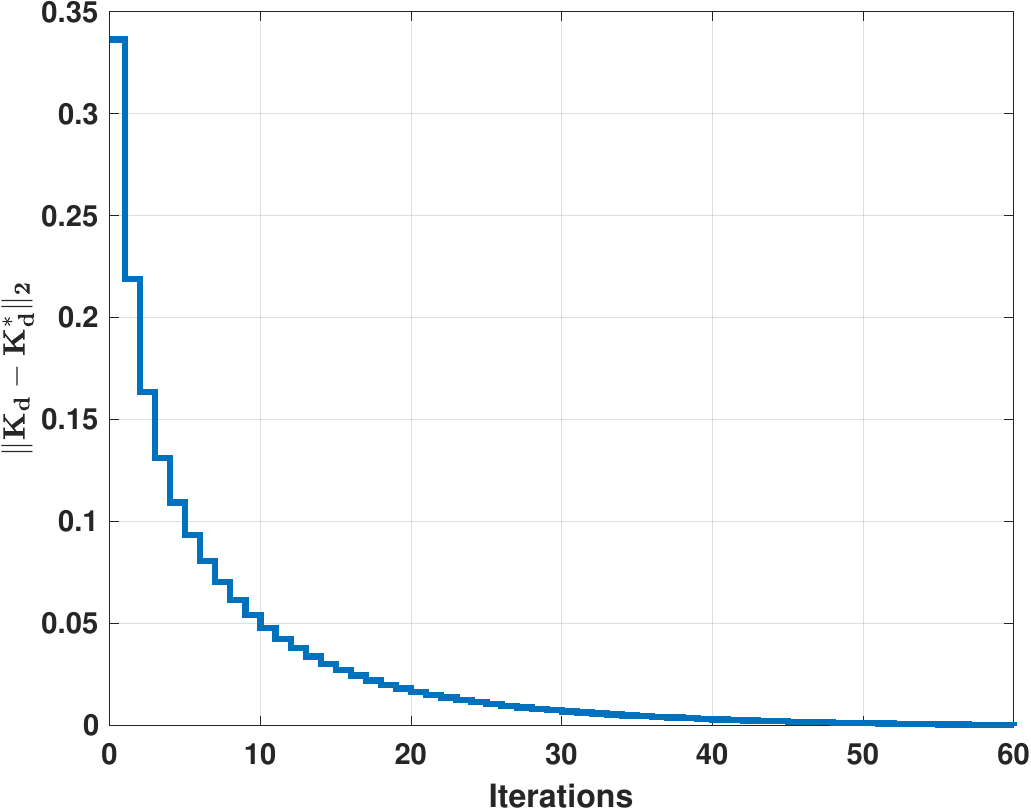}
 \caption{Convergence of the disturbance action network parameters ($K_d$).}
  \label{fig: norm Kd 0_60}
  \end{subfigure}
  	\caption{Convergence of the parameters of actions and critic network.
   % (a) Online convergence of $P$; (b) Convergence of the control action networks parameters ($K_v$).; (c) Convergence of the disturbance action network parameters ($K_d$).
   }
  	  \label{fig: Convergence}
  	\end{figure}
The parameters of the critic and the actions networks are initialized to identity and zero, respectively. Based on this initialization step, the system dynamics move forward in time, and tuning the parameter structures is done by observing the states online. In the RLS problems, the persistency of the excitation condition required to converge the recursive least-squares tuning, i.e., avoiding the parameter drift problem, will hold. However, In Algorithm \ref{Alg: Alg1}, the persistency of the excitation condition is only required for the initial iteration.

In the studied network depicted in Fig. \ref{fig:Campus zones}, $m_1 = 66$, $m_2 = 60$, and $m_3 = 30$. Let’s denote $F_{\rm npe}$ the number of parameters estimated in the matrix $F$. So, $S_{\rm npe}=12246$, $P_{\rm npe}=2211$, ${K_v,}_{\rm npe} = 3960$, and ${K_d,}_{\rm npe} = 1800$.

In Fig. \ref{fig: norm P 0_60}, the convergence of the critic network is illustrated. Fig. \ref{fig: norm Kc 0_60} shows the convergence of the control action network, while Fig. \ref{fig: norm Kd 0_60} depicts the convergence of the disturbance action network.
% the different weight matrix $R_x$, $P$ converge to the optimal value as shown in Fig. \ref{fig: Value Function Param}.

% Table~\ref{tab:tableEQ} shows the average queue length and the average number of rebalancing and customer-carrying vehicles. The metrics are pretty similar for both model predictive control (a model-based approach) and the proposed RL algorithm.

% In Fig. \ref{fig: k_K_opt}, $\left| K-K_{optimal}\right|_2$ converges to zero, and consequently, the action parameters are converging to the optimal values.

\begin{figure}[thpb]
    \centering
  \includegraphics[width=.49\textwidth]{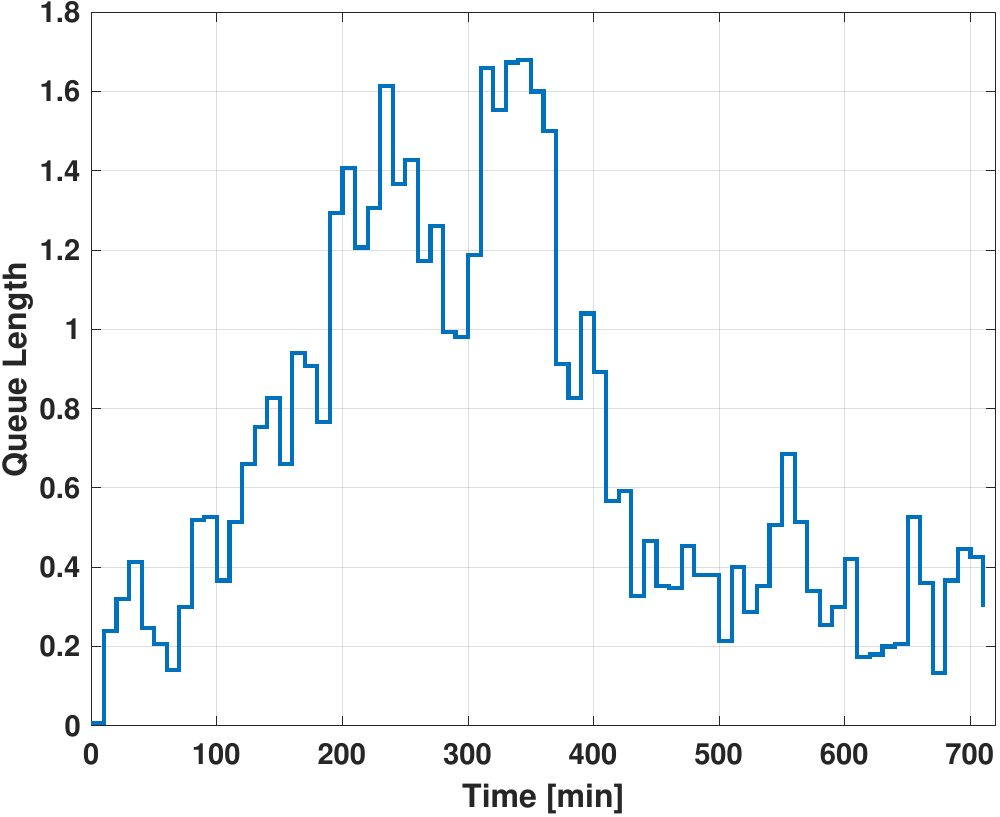}
 \caption{The average queue length.}
  \label{fig: norm Q_agreggate 0_360} 
\end{figure}

\begin{figure}[thpb]
    \centering
  \includegraphics[width=.49\textwidth]{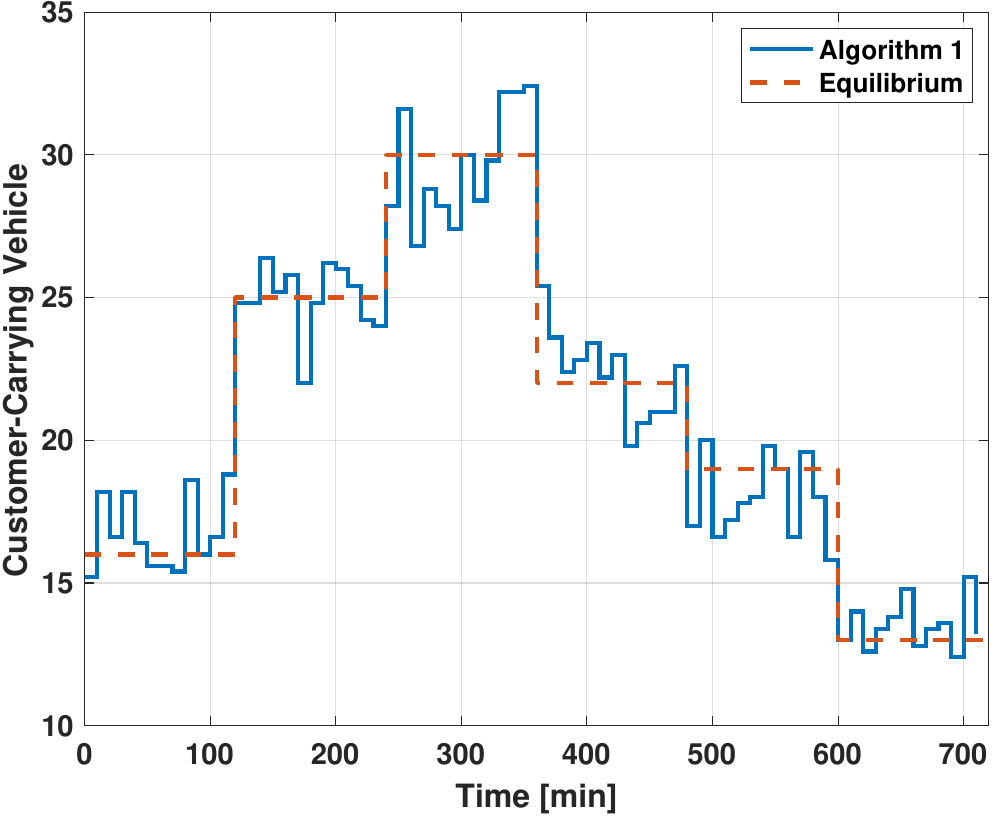}
 \caption{The average customer-carrying vehicles.}
  \label{fig: U_agreggate 0_360}
\end{figure}

\begin{figure}[thpb]
    \centering
  \includegraphics[width=.49\textwidth]{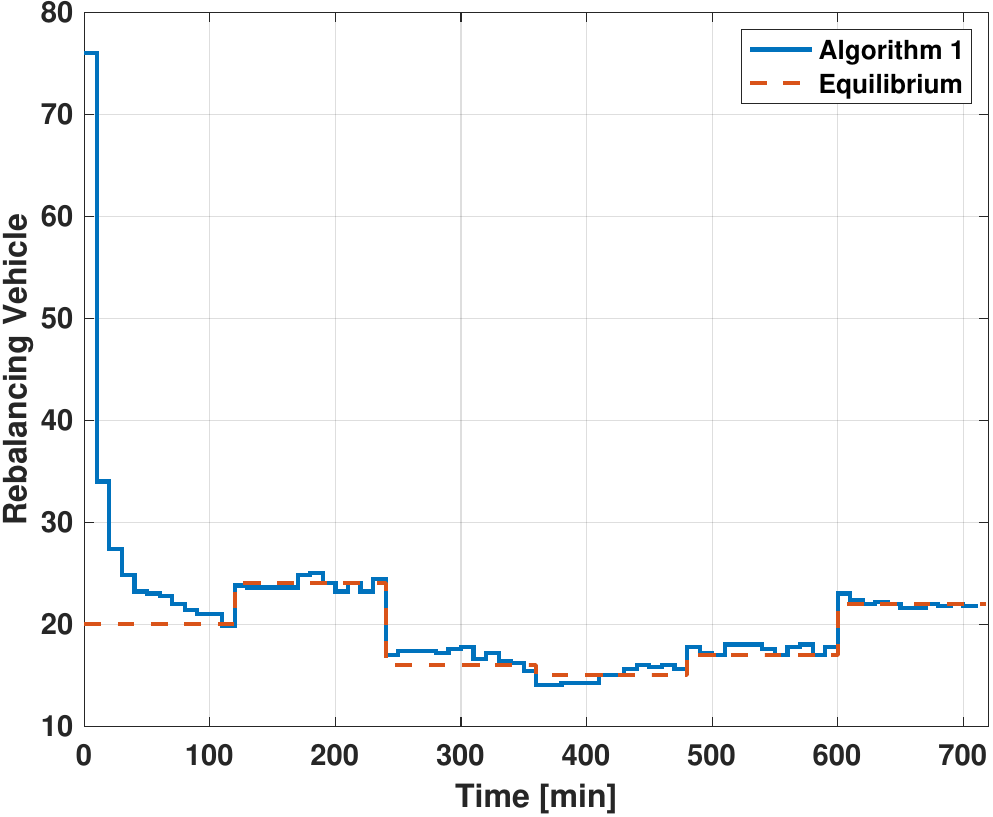}
 \caption{The average rebalancing vehicles.}
  \label{fig: R_agreggate 0_360}
\end{figure}

Figure \ref{fig: norm Q_agreggate 0_360} shows the average queue length over all origin-destinations. At the beginning of the peak hours, the queue length increases while after peak hours, it decreases remarkably. Note that each step of the trace in Figures \ref{fig: norm Q_agreggate 0_360}, \ref{fig: U_agreggate 0_360}, and \ref{fig: R_agreggate 0_360} represents ten minutes.  

Figure \ref{fig: U_agreggate 0_360} depicts the average customer-carrying vehicles in the network. The dashed (red) figure, shows the average of $ \lambda$ over all origin-destinations. Note that we showed in Section \ref{sec: AMoD system modeling} that $\bar{U} = \lambda$. Since the arrival of customers  $d_t$ is given by the realization of a Poisson process of parameter $\lambda$, the expectation of the average customer-carrying vehicles on the links in the network tracks the average of $\lambda$.

Figure \ref{fig: R_agreggate 0_360} illustrates the average number of rebalancing vehicles over all origin-destinations. The dashed (red) figure, shows the average of ${{\bar R}^{\star}}$ over all origin-destinations obtained by \eqref{OptForRebalance}. Similar to Fig. \ref{fig: U_agreggate 0_360}, since the arrival of customers $d_t$ is not deterministic, the expectation of the average number of rebalancing vehicles on the links in the network tracks the average of ${{\bar R}^{\star}}$. Moreover, the optimal rebalancing policy of Algorithm \ref{Alg: Alg1} changes over time due to the change in the demand during the different time intervals.

\section{Conclusion}
\label{sec: conclusion}
In this paper, we proposed a model-free, real-time, data-efficient Q-learning-based algorithm to solve the H$_{\infty}$ control of linear discrete-time systems and applied it to AMoD systems modeled as a H$_{\infty}$ control of the linear discrete-time system.

Besides presenting the algorithm, its convergence was discussed and proved. Afterward, we showed the parameters of the actions and critic networks converged to the optimal values. Numerical results from an AMoD system control in a real case study showed that the proposed algorithm can be implemented in high-dimension systems thanks to the quadratic computational complexity $\mathcal{O}({\underline q}^2)$ and using only a single data point for updating the actor and critic networks in each iteration. 
\section*{Acknowledgement}
% % \section{Acknowledgement}
This research is conducted at the University of Minnesota Transit Lab, currently supported by the following, but not limited to, projects:
\begin{itemize}
     \item[--] National Science Foundation, award CMMI-1831140
     \item[--] Freight Mobility Research Institute (FMRI), Tier 1 Transportation Center, U.S. Department of Transportation%: award RR-K78/FAU SP\#16-532 AM2 and AM3
     \item[--] Minnesota Department of Transportation%, Contract No. 1003325 Work Order No. 44 and 111 
\end{itemize}

\bibliographystyle{ieeetr}
\bibliography{main}

\end{document}